%% file: main.tex
\documentclass[sigconf,authorversion,nonacm]{acmart} 
 \renewcommand\footnotetextcopyrightpermission[1]{}

\usepackage{subcaption}
\AtBeginDocument{%
  \providecommand\BibTeX{{%
    \normalfont B\kern-0.5em{\scshape i\kern-0.25em b}\kern-0.8em\TeX}}}

\usepackage{amsmath}
\usepackage{color}
\usepackage{stfloats}
\usepackage{multirow}
\usepackage{cuted}
\usepackage{url}
\begin{document}

\title{Individualizing Head-Related Transfer Functions for Binaural Acoustic Applications}

\author{Navid~H.~Zandi}
\affiliation{%
  \institution{McMaster University}
  \city{Hamilton}
  \country{Canada}}
\email{hossen6@mcmaster.ca}

\author{Awny~M.~El-Mohandes}
\authornote{This research was done when the author was a Postdoctoral Fellow at McMaster University}
\affiliation{%
  \institution{Mansoura University}
  \city{Mansoura}
  \country{Egypt}}
\email{awny.elmohandes@mans.edu.eg}

\author{Rong~Zheng}
\affiliation{%
  \institution{McMaster University}
  \city{Hamilton}
  \country{Canada}}
\email{rzheng@mcmaster.ca}

\begin{abstract}
A Head Related Transfer Function (HRTF) characterizes how a human ear receives sounds from a point in space, and depends on the shapes of one’s head, pinna, and torso. Accurate estimations of HRTFs for human subjects are crucial in enabling binaural acoustic applications such as sound localization and 3D sound spatialization. Unfortunately, conventional approaches for HRTF estimation rely on specialized devices or lengthy measurement processes. This work proposes a novel lightweight method for HRTF individualization that can be implemented using commercial-off-the-shelf components and performed by average users in home settings. The proposed method has two key components: a generative neural network model that can be individualized to predict HRTFs of new subjects from sparse measurements, and a lightweight measurement procedure that collects HRTF data from spatial locations. Extensive experiments using a public dataset and in house measurement data from 10 subjects of different ages and genders, show that the individualized models significantly outperform a baseline model in the accuracy of predicted HRTFs. To further demonstrate the advantages of individualized HRTFs, we implement two prototype applications for binaural localization and acoustic spatialization. We find that the performance of a localization model is improved by 15$^\circ$ after trained with individualized HRTFs. Furthermore, in hearing tests, the success rate of correctly identifying the azimuth direction of incoming sounds increases by 183$\%$ after individualization.
\end{abstract}


\maketitle

\section{Introduction}\label{sec:introduction}
With their explosive adoption and rich functionalities, earable devices (or {\it earables}) are becoming the new frontier of mobile computing. Researchers have explored in recent years the use of embedded sensors on such devices for biosensing such as body temperature monitoring\cite{ns01}, heart rate monitoring \cite{ns02}, and biometric-based user authentication ~\cite{ns02, ns04}, or as an extra input for gesture-based control \cite{ns05}. Additionally, earables can also bring better environmental awareness and immersive acoustic experiences to users in augmented reality (AR) and virtual reality (VR) applications. For example, with binaural sound localization and recognition, a user can be alerted of imminent dangers in one's surroundings (e.g., an approaching vehicle) while on a call or listening to loud music; with 3D acoustic spatialization, playbacks of sounds through in-ear speakers are customized to make users feel that they are actually in a 3D environment. To enable these binaural acoustic applications on earables, an in-depth understanding of how human auditory systems perceive and process acoustic events spatially is essential.

Humans can perceive the direction of incoming sounds. Even in a cluttered environment, like in a restaurant or a stadium, one is capable of separating and attending to individual sound sources selectively \cite{wang2006computational}. Our ability to localize sound is attributed to the filtering effects of the ear, head and torso, which are direction and frequency dependent, and are described by Head-Related Transfer Function (HRTF) \cite{iida2019head}. HRTF characterizes the way sounds from different points in space are perceived by the ears, or in other words, a transfer function of the channel between a sound source and the ears. Consequently, HRTF is a function of the angles of an incoming sound (e.g., azimuth and elevation angles in 3D interaural coordinates) and frequency, and is defined separately for each ear.

\begin{figure*}[!t]
\centering
\subfloat[]{\includegraphics[width=1.8in]{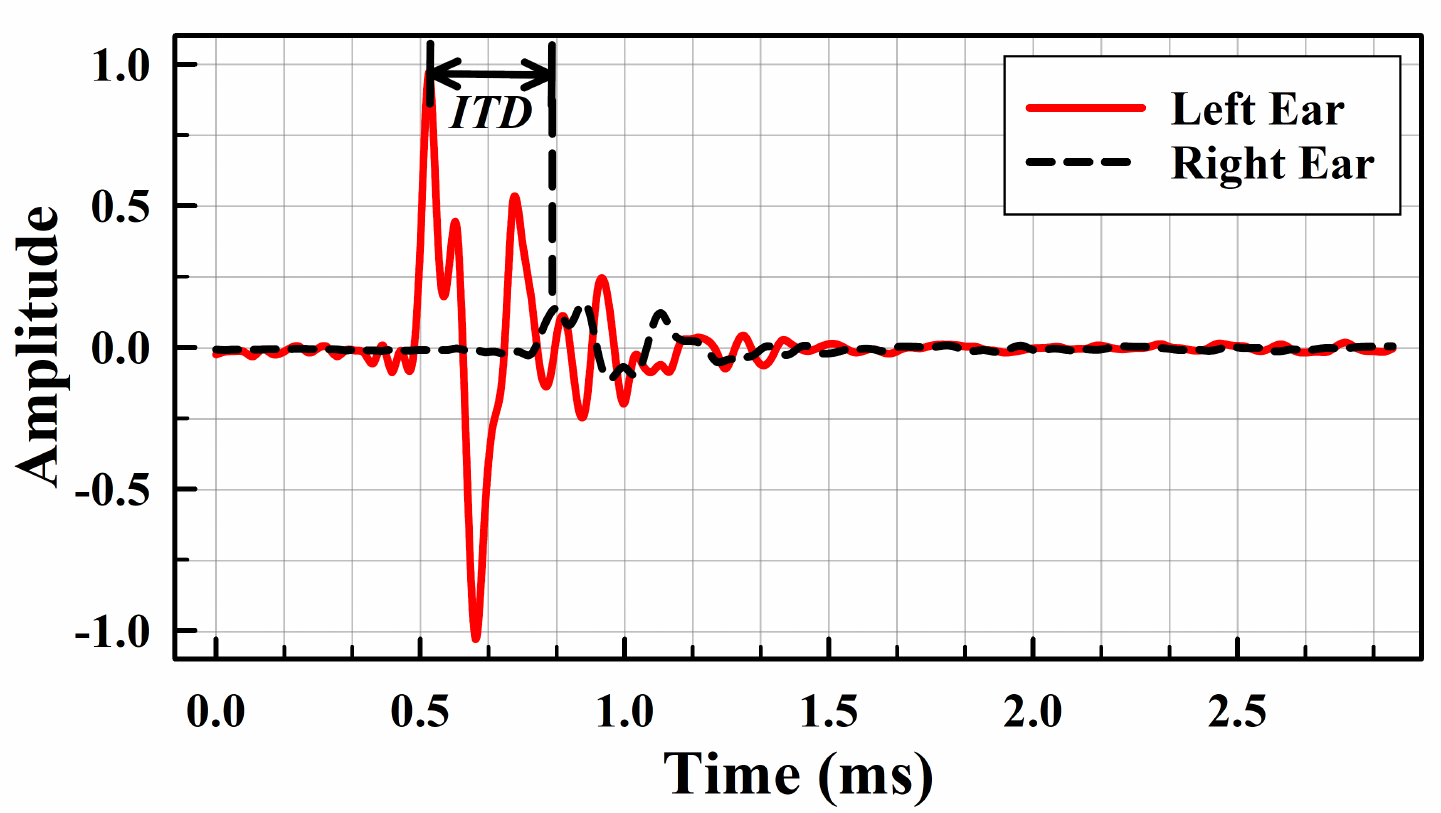}%
\label{fig:intrd_hrir}}
\hfil
\subfloat[]{\includegraphics[width=1.8in]{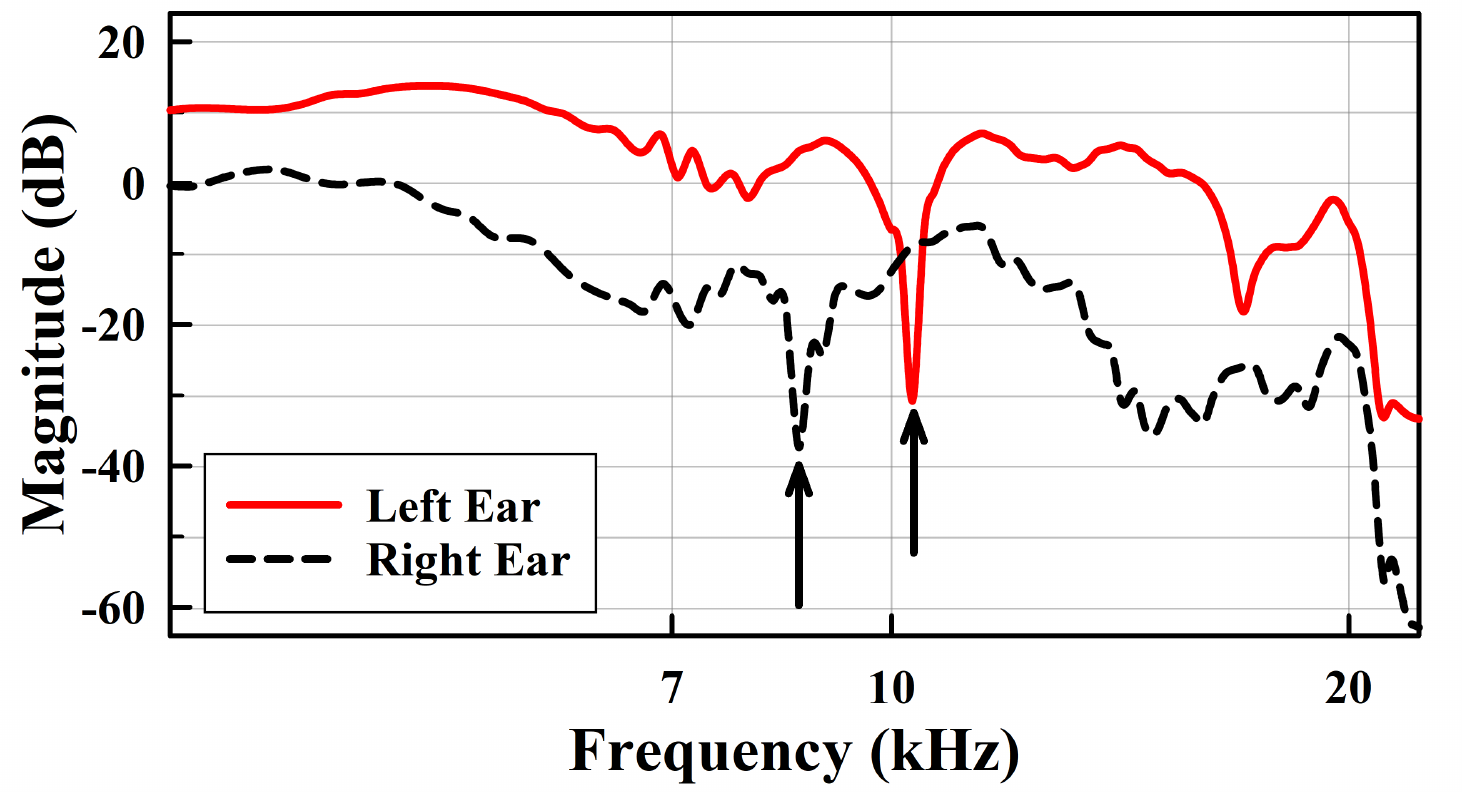}%
\label{fig:intrd_hrtf_1}}
\hfil
\subfloat[]{\includegraphics[width=1.8in]{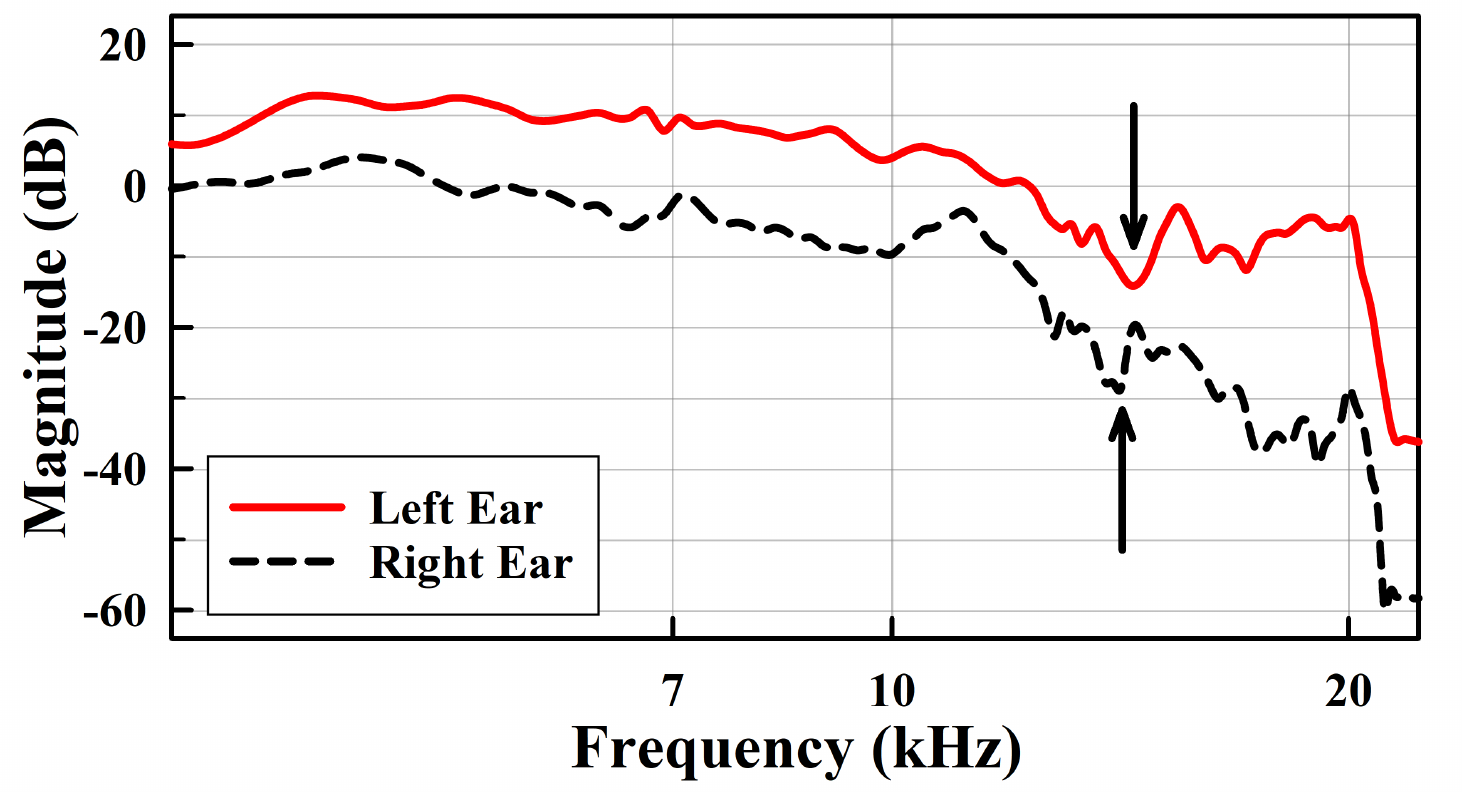}%
\label{fig:intrd_hrtf_2}}
\caption{Examples of HRIRs and HRTFs. (a) HRIR at azimuth = $-45^{\circ}$ and elevation = $-5.76^{\circ}$. (b) HRTF at azimuth = $-45^{\circ}$ and elevation = $-5.7^{\circ}$, and (c) HRTF at azimuth = $45^{\circ}$ and elevation = $54.72^{\circ}$ (some notches are marked with arrows). The notches appear at higher frequencies as the position of the sound moves toward the top of one's head.}
\label{fig:three graphs}
\vspace{-1em}
\end{figure*}

An example of HRTF and its counterpart in the time domain (HRIR) is given in Figure \ref{fig:three graphs}, for left and right ears. As shown in Figure \ref{fig:intrd_hrtf_1}, many peaks and notches can be observed. As the position of the sound source goes toward the top of the head, the frequencies of spectral notches become higher (Figure \ref{fig:intrd_hrtf_2}). The perception of the elevation angle of a sound is related to the spectral notches and peaks above 5kHz \cite{hebrank1974spectral}. These spectral cues depend on the direction of the incoming signal as well as human physical features, such as the shapes and sizes of one’s pinna, head, and torso. On the other hand, time difference (ITD) and level difference (ILD) between the sounds received by the left and right ears are the two main cues for lateral localization, and they are directly affected by human’s HRTF \cite{iida2019head}. Since HRTFs are highly specific to each person, using another person’s HRTF or a generic one will lead to errors in acoustic localization and unpleasant experiences in audio playbacks for humans. However, since HRTFs depend on the location of the sound, direct measurements are time-consuming and generally require special equipment. Developing efficient mechanisms to estimate subject-specific HRTFs, also called \emph{HRTF individualization}, to enable biaural acoustic applications has been an active area of research in recent years.

In this work, we propose a novel lightweight algorithm for HRTFs individualization, which with the help of a simple measurement procedure, makes it possible to individualize HRTFs at home using mobile devices. This measurement procedure is fast, lightweight and easy to conduct using commercial-off-the-shelf (COTS) components, and can result in accurate HRTFs for target users to facilitate the aforementioned applications. Our algorithm is based on a \emph{conditional variational autoencoder (CVAE)} to learn the latent space representation of input data. Given measurement data from sparse positions, the model can be adapted to generate individualized HRTFs for all directions. Compared to a state-of-the-art model in \cite{CVAE-02}, the proposed CVAE model has 470 times fewer parameters, making it attractive for implementation on embedded devices. Moreover, it has higher accuracy in representing the characteristics of both training and test subjects. After training the model on a public HRTF dataset, the HRTFs of a new user can be accurately estimated using measurements from as low as 60 locations from the user. The proposed lightweight measurement procedure uses two microphones to record sounds emitted from a mobile phone. Positions of the phone are estimated from on-board inertial measurement units (IMUs) in a global coordinate frame and transformed to a subject-specific frame. No anthropometric information is required from users. The total measurement time is less than $5$ minutes according to our user study, significantly shorter than the procedure reported in \cite{CVAE-02}.

Extensive evaluations have been conducted using both the public dataset and real measurement data. We find that the adapted model improves HRTF prediction accuracy by $31\%$ and $39\%$ on average in leave-one-subject-one experiments compared to non-adapted model. Interestingly, we find that by using only sparse measurements from the half sphere in front of one’s body, the errors only increase by $8\%$. The proposed measurement procedure has a mean angle of arrival estimation error of $4.7^\circ$. To further demonstrate the advantage of individualized HRTFs, we have implemented two applications: binaural localization and acoustic spatialization. We find that the average error of a subject-independent localization model is reduced by $15.73^\circ$ after adaptation. For sound spatialization, audio signals at $24$ different positions are generated by convoluting a mono sound with HRTFs with and without individualization for $10$ volunteers separately. We find that the users can detect azimuth angles correctly $82.55\%$ of the time compared to $29.17\%$ of the time without individualization.

The rest of this paper is organized as follows. The state-of-the-art methods for HRTF individualization are discussed in Section 2. In Section 3 and 4, the algorithm overview and the neural network architecture of the proposed CVAE model are described. In Section 5, we present the details of the measurement procedure. The performance of the proposed model is evaluated using a public dataset and data collected through our measurement procedure in Section 6. Finally, in Section 7, we present the implementation and evaluation results of two main applications of HRTF, localization and sound spatialization, followed by a discussion of limitations of the work and conclusion in Section 8.

\section{Related Works}\label{Related_work}
\subsection{HRTF Individualization}
Using generic HRTFs is the main source of errors in many applications. Existing approaches to individualize HRTFs can be grouped into four main categories \cite{Guezenoc-01}, as discussed in this section.

\subsubsection{Direct Methods}
The most obvious solution to obtaining individualized HRTFs of a subject is to conduct dense acoustic measurements in an anechoic chamber \cite{ITA-01}. Several loud speakers are positioned surrounding the subject in all directions of interest with microphones placed at the entrance of ear canals recording the impulse responses. The number of required speakers can be reduced by installing them at different elevations on an arc, and rotating the arc to measure at different azimuths. This approach requires special devices and setups. The measurement procedure can be overwhelming to test subjects (often having to sit still for a long time). To accelerate the process, the Multiple Exponential Sweep Method (MESM) is employed in \cite{majdak2007multiple}, where reference signals are overlapped in time. However, this method requires a careful selection of timing to prevent superposition of different impulse responses. An alternative way is the so-called reciprocal method \cite{zotkin2006fast}, in which two small speakers are placed inside the subject's ears, and microphones are installed on an arc. This accelerates the measurement time, but has its own limitations, as the speakers in the ears can not produce too loud sounds as it may damage the person's ears (low SNR on the final measurements). Recently, some researchers investigate the use of continuous measurements. In \cite{richter2019influence}, using measurement in an anechoic room, it is reported that at a rotation speed of $3.8^\circ/s$, no audible differences are experienced by subjects compared to step-wise measurement. In \cite{reijniers2020hrtf}, instead of moving her whole body, a subject is asked to move her head in different directions, with the head movements tracked by a motion tracker system. Long measurement time often leads to motion artifacts due to subject movements during the measurements \cite{hirahara2010head}. Some have attempted to alleviate this problem by providing visual feedback to subjects \cite{denk2017controlling}, or constraining their movements by mechanical supports.
Another approach to ease the need for special measurement setups is proposed in \cite{yang2021personalizing}. A user measures her HRTFs at 2D sparse positions in a horizontal plane using a mobile phone. HRTF estimates are done by modeling sound diffraction on the head, with the help of the physics of sound propagation fused with mobile IMU data to locate the mobile phone's position. To obtain HRTF estimations at unobserved locations, linear interpolation is used. Liner interpolation using neighboring measurements does not work well when the measurement  locations are far apart. Furthermore, since it is difficult to take measurements behind one’s back, extrapolation based on frontal measurements leads to large errors. For example, as reported in \cite{yang2021personalizing}, two out of 5 volunteers had difficulty to correctly measure HRFTs in their back due to arm movement constraints. The estimation is limited to HRTFs in a horizontal plane for users, and the effect of HRTF individualization is not tested on sound spatialization experience. In contrast, our method targets 3D HRTF estimation in both azimuth and elevation. The proposed model learns representations of the full sphere from the ITA datasets. Even in the absence of measurement data in the back during adaptation, the model can still leverage the learned representation to generate HRTFs in those positions. 

\subsubsection{Simulation-based Methods}
The second categories of HRTF individualization methods utilize numerical simulations of acoustic propagation around target subjects. To do so, a 3D geometric model of a listener's ears, head, and torso is needed either through 3D scans or 3D reconstruction from 2D images \cite{kaneko2016deepearnet}. Methods like finite difference time domain \cite{mokhtari2007comparison}, boundary element \cite{gumerov2007fast}, finite element \cite{huttunen2007simulation}, and raytracing \cite{rober2006hrtf} are employed in numerical simulations of HRTFs. The accuracy of the 3D geometric model as inputs to these simulations is key to the accuracy of the resulting HRTFs. In particular, ears should be modeled more accurately than the rest of the body. Objective studies have reported good agreement between the computed HRTFs in simulation-based methods, and those from fine-grained acoustic measurements \cite{gumerov2007fast}. Numerical simulations tend to be compute intensive, but thanks to the ever-growing computation power, and improved algorithms, such simulations can be completed under an hour for one subject \cite{meshram2014efficient}. However, most methods still require special equipment such as MRI or CT for 3D scan, and are thus not accessible to everyone. 3D reconstruction from 2D images eliminate the need for specialize equipment but at the expense of lower accuracy.

\subsubsection{Indirect Methods Using Anthropometric Measurements}
HRTFs rely on the morphology of the listener. Therefore, many works have tried to indirectly estimate HRTFs from anthropometric measurements. Existing methods can be further classified into three subcategories.

\noindent \textbf{Adaptation:} Starting from a non-individualized HRTF, scaling in the frequency domain can be applied for individualization \cite{middlebrooks1999individual}, where the scaling factors can be estimated from head and pinna measurements \cite{middlebrooks2000psychophysical}. Subjective evaluations on 9 to 11 subjects showed improved localization performance over non-individualized HRTFs. Further improvement can be achieved by combining frequency scaling with rotation in space to compensate for head tilt \cite{maki2005reducing}.

\noindent \textbf{Nearest neighbor selection:} In these approaches, the nearest HRTF set in a dataset is first selected based on the anthropometric measurements. The distances between two subjects can be computed either directly from morphological parameters \cite{zotkin2002customizable}, or features output from a neural network \cite{shu2017head}. Adaption can be further applied using methods in the previous category.

\noindent \textbf{Regression:} The third category of approaches try to establish a functional or stochastic relation between anthropometric parameters and characteristics parameters of HRTFs. Principle component analysis (PCA) are often used to reduce the dimensionality of input and/or output parameters \cite{hu2006head}. In \cite{hu2006head}, a linear model is assumed and the HRTFs for a new subject is predicted using the subject's  anthropometric parameters through the model. Good agreements are reported at different azimuth angles in subjective evaluation, but elevation angles performance is not studied. \cite{lee2018personalized} and \cite{chen2019autoencoding} extend the above work by modeling the two set of parameters using a deep neural network and an autoencoder network, respectively.

All methods in this category suffer the same problem as simulation-based methods in their needs for accurate anthropometric measurements, which are often difficult to obtain.

\subsubsection{Indirect Methods based on Perceptual Feedback}
Beside using anthropometric parameters to identify closely matched subjects in a dataset, a fourth category of approaches utilizes perceptual feedback from target listeners. A reference sound which contains all the frequency ranges (Gaussian noise, or parts of a music) is convoluted with selected HRTFs in a dataset and played through a headphone to create 3D audio effects. The listener then rates among these playbacks how close the perceived location of the sound is to the ground truth locations. Once the closest $K$-subjects in the dataset are found, the final HRTF of the listener can be determined through selection or adaptation. 

In {\it selection} methods, the closest non-individualized HRTFs from the dataset are used \cite{katz2012perceptually}. The reported tuning time ranges from 15 minutes to more than 35 minutes. In contrast, {\it adaptation} uses frequency scaling with scaling factors tuned by the listener's perceptual feedback \cite{holzl2014global,CVAE-02}. In \cite{CVAE-02}, the authors first train a conditional variational autoencoder to generate HRTFs using a public dataset. During adaption, user feedback is collected and is used to optimize personalized weights of HRTFs from known subjects in the dataset. It reports significant improvement over the non-individualized HRTF for 18 out of 20 subjects. However, collecting user feedback requires minimally 30 minutes.

Methods using perceptual feedback generally suffer from long calibration time and imperfection of human hearing (e.g, low resolutions in elevation angles, difficulty to discriminate sounds in front or behind one's body). 
Our proposed method is a combination of the direct and indirect methods. It uses HRTF estimations at sparse locations from a target subject (direct measurements), and estimates the full HRTFs with the help of a latent representation of HRTFs using a deep generative model (indirect adaptation). To the best of our knowledge, this is the first work to combine both lines of approaches.

\subsection{HRTF Datasets}
Several datasets are available for HRTF measurements using anechoic chambers. They differ in the number of subjects in the dataset, the spatial resolution of measurements, and sampling rates. The CIPIC dataset \cite{algazi2001cipic} contains data from 45 subjects. With a spacing of $5.625^{\circ}\times 5^{\circ}$, measurements were taken at 1250 positions for each subject. A set of 27 anthropometric measurements of head, torso and pinna are included for only 43 subjects. The LISTEN dataset \cite{LISTEN} measured HRTFs of 51 subjects at 187 positions recorded with a resolution of $15^{\circ} \times 15^{\circ}$. The anthropometric measurements of the subjects are also included. A larger dataset, RIEC \cite{watanabe2014dataset, RIEC}, contains HRTFs of 105 subjects with a spatial resolution of $5^{\circ}\times10^{\circ}$, totaling 865 positions. A 3D model of head and shoulders is provided for 37 subjects. ARI \cite{ARI} is the largest HRTF dataset with over 120 subjects. It has a resolution of $5^{\circ}\times5^{\circ}$, with $2.5^{\circ}$ horizontal steps in the frontal space. For 50 of the subjects, a total of 54 anthropometric measurements are available, out of which 27 measures are the same as those in the CIPIC dataset. The ITA dataset \cite{ITA-01} has a high resolution of $5^{\circ}\times5^{\circ}$, with a total of 2304 HRTFs measured for 48 subjects. Using Magnetic Resonance Imaging (MRI), detailed pinna models of all the subjects are available.

\section{Solution Overview}

\begin{figure}[!t]
    \centering
    \includegraphics[width=3.5in]{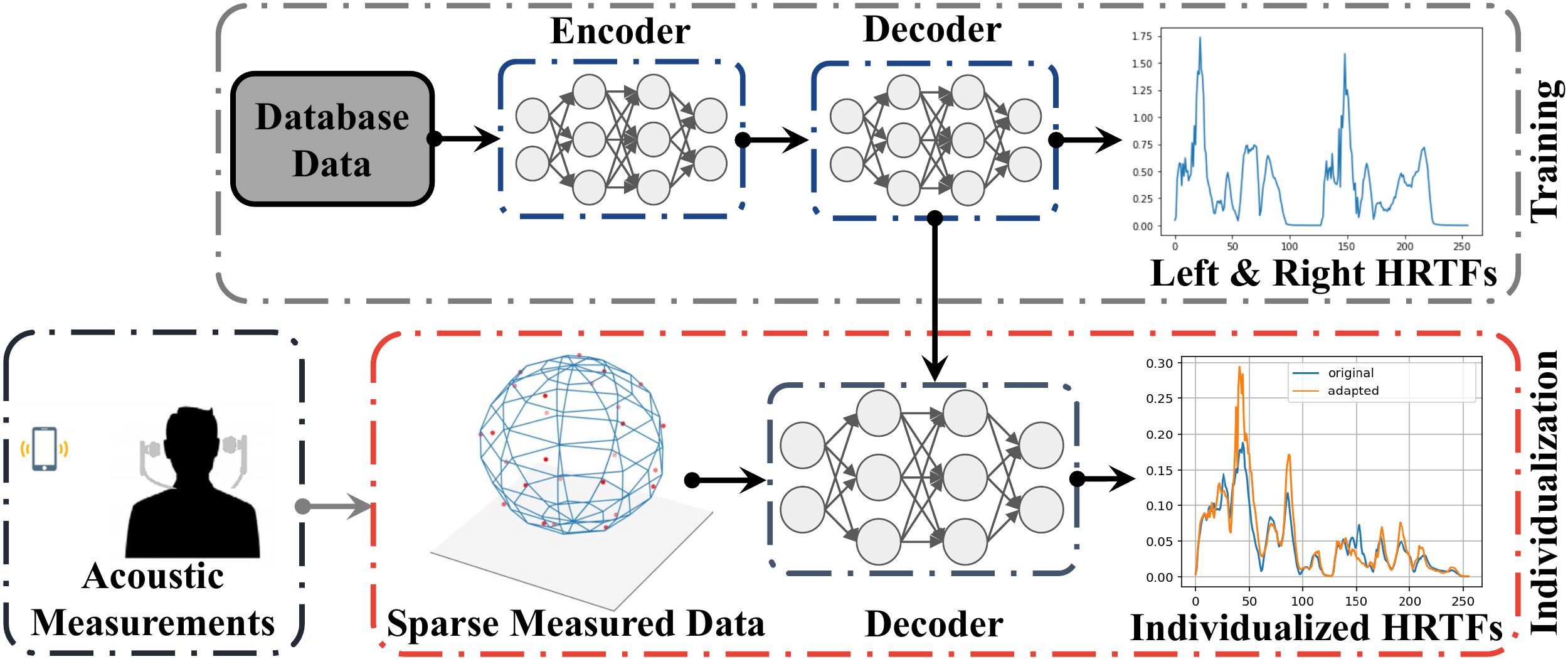}
    \caption{System diagram of the proposed approach}
    \label{fig:diagram}
\vspace{-1em}
\end{figure}

As shown in Figure \ref{fig:diagram}, the proposed HRTF individualization approach consists of three main components: 1) training a conditional variational autoencoder (CVAE) using HTRF data from existing datasets, 2) collecting sparse measurements from a target subject using COTS devices, and 3) individualization for the new subject based on sparse measurements. Specifically, we first train a CVAE network using data from 48 subjects in the ITA HRTF dataset, to learn the latent space representation for HRTFs at different positions in the space. The network takes as inputs HRTFs from the left and right ears, the direction of the HRTFs, and a one-hot encoded subject vector. After training, the decoder in the CVAE model can generate HRTFs for any subject in the dataset at arbitrary directions by specifying the subject index and direction vectors as inputs. However, it cannot be used to generate HRTFs for a {\it specific} subject not part of the training dataset. To obtain individualized HRTFs, we need to first collect some measurement data from the target subject.

\begin{figure}[!t]
    \centering
    \includegraphics[width=2.73in]{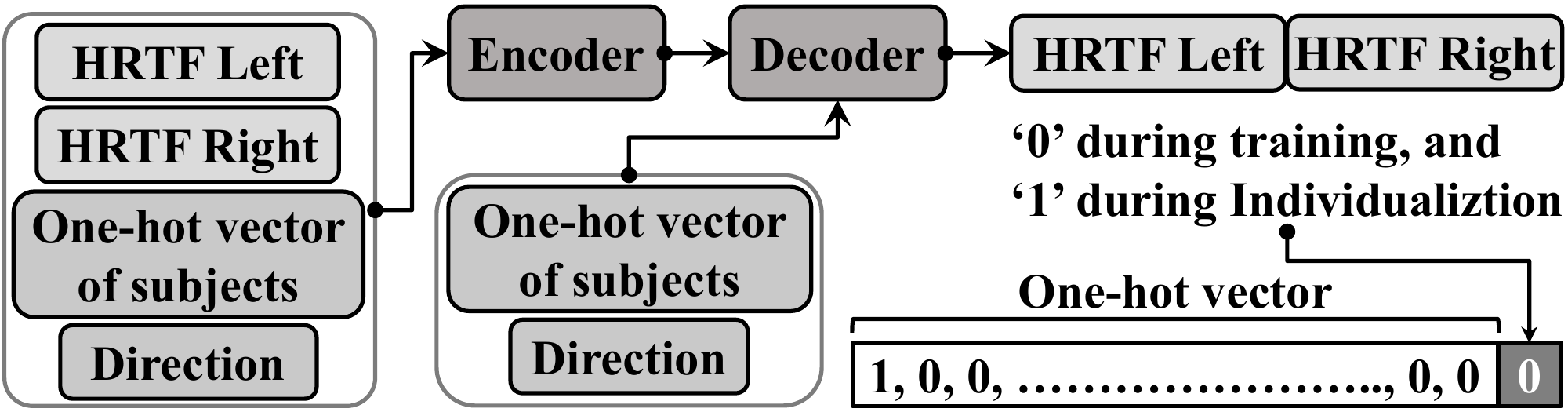}
    \caption{The input and output of the proposed CVAE model}
    \label{fig:CVAE}
\vspace{-1em}
\end{figure}

This is accomplished by the data collection component, where a subject puts on a headset and records the sounds impinging upon the in-ear microphones from a reference signal emitted from a mobile phone. The subject freely moves the mobile phone with her right and left hands in the space. To tag estimated HRFTs with appropriate location labels, we need to determine the relative position of the mobile device to the subject. This is non-trivial without the knowledge of anthropometric parameters of the subject. However, we devise a sensor fusion mechanism to transform device poses from the device frame to the body frame of the subject. Finally, the positionally labeled data will be used in adapting the decoder of the CVAE to generate individualized HRTFs for the subject at arbitrary directions. In subsequent sections, we will present  the three components in details.

\section{A Generative Model for HRTFs}
Recently, deep generative models have been widely used to generate highly realistic data such as images and music. In this paper, we utilize a CVAE to propose a lightweight generative model, that is suitable for mobile devices and embedded systems computational resources. The proposed model estimates the subject-specific HRTFs using a sparse measured data from this subject.

\subsection{Network Architecture \label{4_network_arch}}
\begin{figure}[!t]
\centering
\subfloat[Encoder]{\includegraphics[width=0.23\textwidth]{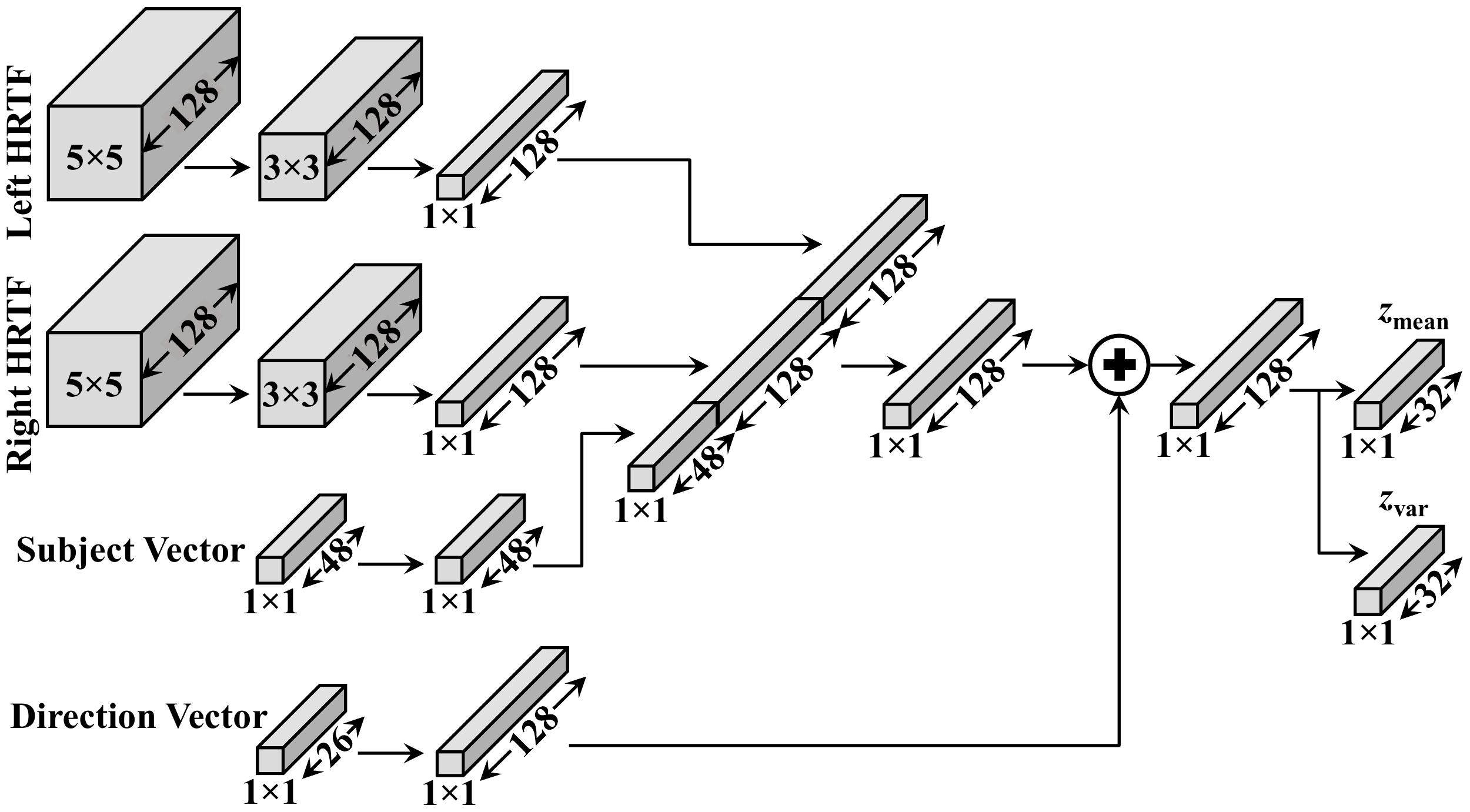}%
\label{fig:Enc_CVAE}}
\hfil
\subfloat[Decoder]{\includegraphics[width=0.23\textwidth]{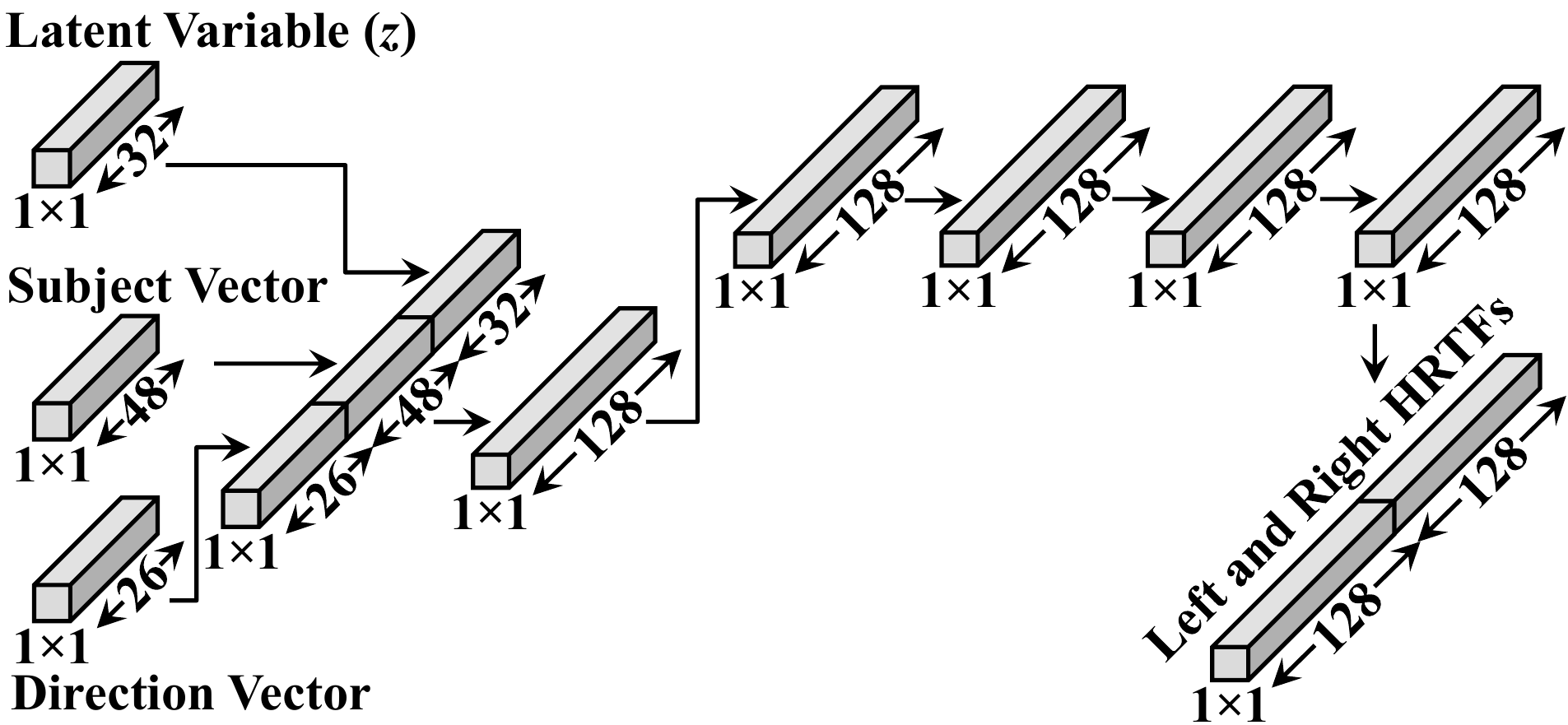}%
\label{fig:Dec_CVAE}}
\caption{Architecture of the CVAE model: (a) the encoder encodes an input HRTF into a latent space representation and (b) the decoder reconstructs the input HRTF based on its direction and subject vector.}
\label{fig:Encoder CVAE}
\end{figure}

The proposed CVAE model consists of an encoder network and a decoder network (Figure \ref{fig:CVAE}). The encoder network (Figure \ref{fig:Enc_CVAE}) takes three inputs: the HRTFs magnitude of a subject, a one-hot vector for the subject, and a direction vector. First, for each subject and each direction in the dataset, we take the subject's HRTFs from $5\times 5$ grid points centred at the desired direction. The grid points are evenly spaced according to Spherical Harmonics representation \cite{SPH-01} with $\pm 0.08\pi$ step size in azimuth and elevation angles. The magnitude of the HRTF is computed at each grid point over 128 frequency bins, resulting a $5\times 5\times 128$ tensor for each ear. The two tensors will separately go through two 3D Convolutional layers with kernel size 3x3x3, stride 1 for all directions, and zero padding only on the last dimension, to form the HRTF's features.

The subject ID is encoded as a one-hot vector. If we have an $N$ subjects in the training set, the vector will be of length $N+1$. The $(N+1)$th element is reserved for the individualization process, discussed next, that is set to zero during the training process. Each subject vector goes through a fully-connected layer, and then concatenated with the output of the CNN layers from the previous step. This concatenated tensor goes through another fully-connected layer and then summed with the output of the fully-connected layer that processes the direction vector of the corresponding HRTF. This vector represents the direction in $\mathbb{R}^{26}$ instead of $\mathbb{R}^3$, where the basis vectors correspond to 26 evenly distributed points on the sphere. 
 
For each desired direction $u$, four enclosing neighboring points \emph{(B1, B2, B3, B4)} are identified, and the weights for the basis vectors $(w_1, w_2, w_3, w_4)$ are calculated accordingly \cite{CVAE-02}:
\begin{equation*} \label{direction}
 (w_1, w_2, w_3, w_4) = (st, (1-s)t, s(1-t), (1-s)(1-t)),
\end{equation*}
where $s = (\phi_u - \phi_{B2})/(\phi_{B1}-\phi_{B2})$, $t = (\theta_u - \theta_{B3})/(\theta_{B1}-\theta_{B3})$, and $\phi$ and $\theta$ are the azimuth and elevation angles of the corresponding points. The weights for directions other than the four surrounding basis vectors are set to zero. Compared to representations in $\mathbb{R}^3$, such a representation is more suitable for processing by neural networks as they are sensitive to binary like activation. Finally, the output of the summation is mapped into a latent variable  $z$, a 1-D vector of size 32.

On the decoder side (Figure \ref{fig:Dec_CVAE}), the latent variable is concatenated with the subject and the direction vectors. The resulted tensor goes through 5 fully-connected layers of the same size (e.g., 128), appended with an output layer to estimate the HRTFs of the specified subject, in the desired direction. Exponential-linear activation functions are used after each layer in the encoder and the decoder, except for the final output layer that uses a sigmoid function. 
The proposed network architecture, though inspired by the CVAE model in \cite{CVAE-02}, has two marked differences. First, we view HRTF generation as a regression problem. The outputs of the decoder are thus floating point vectors of size 256 (128 for each ear). In contrast, the model in \cite{CVAE-02} quantizes the values in each of the 128 frequency bins into 256 levels and outputs a one-hot vector of dimensions $256\times 256$. Doing so drastically increases the number of parameters in the network due to a large number of units in the output layer. Second, no adaption layers are included in our model, which further reduces the number of learning parameters. As a result, the total number of parameters in our model and the one in \cite{CVAE-02} are 133,390 and 637,100,032, respectively. A lower number of training parameters generally implies shorter training time and higher data and computation efficiency.

\subsection{Individualization \label{sec:individualization_decoder}}
After training, the decoder can be used to generate HRTFs at an arbitrary direction for any subject in the training dataset. However, we need to fine-tune and adapt its hyperparameters to generating HRTFs for a new subject. Therefore, we need to measure the new subject's HRTFs at sparse locations, a procedure detailed in Section \ref{real-data}, and use this data to adapt the decoder. 
The decoder model can be adapted using sparse data because of the underlying (sparse) structure of the HRTFs. The locations of peaks and notches in HRTFs depend on the azimuth and elevation angles, and are highly correlated among neighboring angles. Our network learns the sparse representations of HRTFs during the training stage. In this way, with HRTFs at a small number of locations as inputs to the network, the decoder can estimate the remaining HRTFs. For adaptation, we re-train the decoder with the new user's measured data and a random batch of data from existing subjects in the dataset to avoid any over-fitting. In the implementation, we utilize $5\%$ of the data in the ITA dataset or equivalently, 5000 data entries.

Specially, for both the new subject and existing subjects, a latent variable, $z$, is sampled from a normal Gaussian distribution, and together with subject and direction vectors are used to re-train the decoder. As mentioned in Section \ref{4_network_arch}, in the subject vector, all elements are zero, except the last one. The outputs of the decoder, before individualization, can be seen as a set that blends different features from all subjects in the training stage, or roughly HRTFs of an average subject. By fine-tuning the decoder parameters using data from the new subject at sparse directions, the locations and amplitudes of the peaks and notches in HRTFs will be adapted for the new subject, leveraging the structure information that the network has learned from existing subjects. We need the phase information to reconstruct the time domain signals from the adapted frequency domain response. Minimum-Phase reconstruction is used, and then the appropriate time delay (ITD) is added to the reconstructed signals based on the direction \cite{holzl2014global}. The ITD is estimated using the average of all users in the dataset, and then scaling it relatively to the new subject based on the new measurements.

\section{Lightweight Data Collection from New Subjects \label{real-data}}
In this section, we present the procedure to collect data from new subjects for HRTF individualization. Compared to the direct measurement methods discussed in Section \ref{Related_work}, the procedure is fast and easy to perform by average users at home. It does not require specialized devices or anthropometric measurements from users.

\subsection{Measurement Setup and Procedure}
\subsubsection{Devices}
The two required devices in our setup are: 1) two in-ear microphones to record the sounds impinging on the subject's ears. Sounds captured  will be transmitted and stored on a computer for post-processing. 2) a mobile phone to play back sounds on-demand. The mobile phone is equipped with IMU sensors to estimate the location of emitted sounds.

\subsubsection{Procedure}
During measurements, a user needs to put the two microphones in her ears, hold the mobile phone in her hand, and stretch out her arm as far as possible from her body. We require the long edge of the mobile phone to be parallel to the extension of the user's arm. During the entire process, the user remains stationary with only upper limb movements. As the user moves her arm around, she can pause at arbitrary locations and play back a pre-recorded sound using the mobile phone. The pre-recorded sound is chosen to be an exponential sine sweep signal, which allows better separation of nonlinear artifacts caused by acoustic transceivers from useful signals compared to white noise or linear sweep waves \cite{majdak2007multiple}. Once the sound finishes playing, the user can proceed to another location, repeating the steps multiple times. There is no special motion pattern for the arm required, but the user should try to cover as much range as possible while keeping her shoulder at the same location. The procedure is repeated for both hands to have the maximum coverage. At each position that the hand stops, two sources of information are obtained: 1) the recorded sounds in the two microphones, and 2) the position that the reference sound is played from. Using these two information, we then calculate the HRTFs at a specific location, by deconvolving the reference sound from the recorded sounds in both ears. Next, we discuss how to determine the directions of sound sources in the procedure without user anthropometric parameters and specialized equipment.

\begin{figure}[!t]
    \centering
     \includegraphics[width=2.8in]{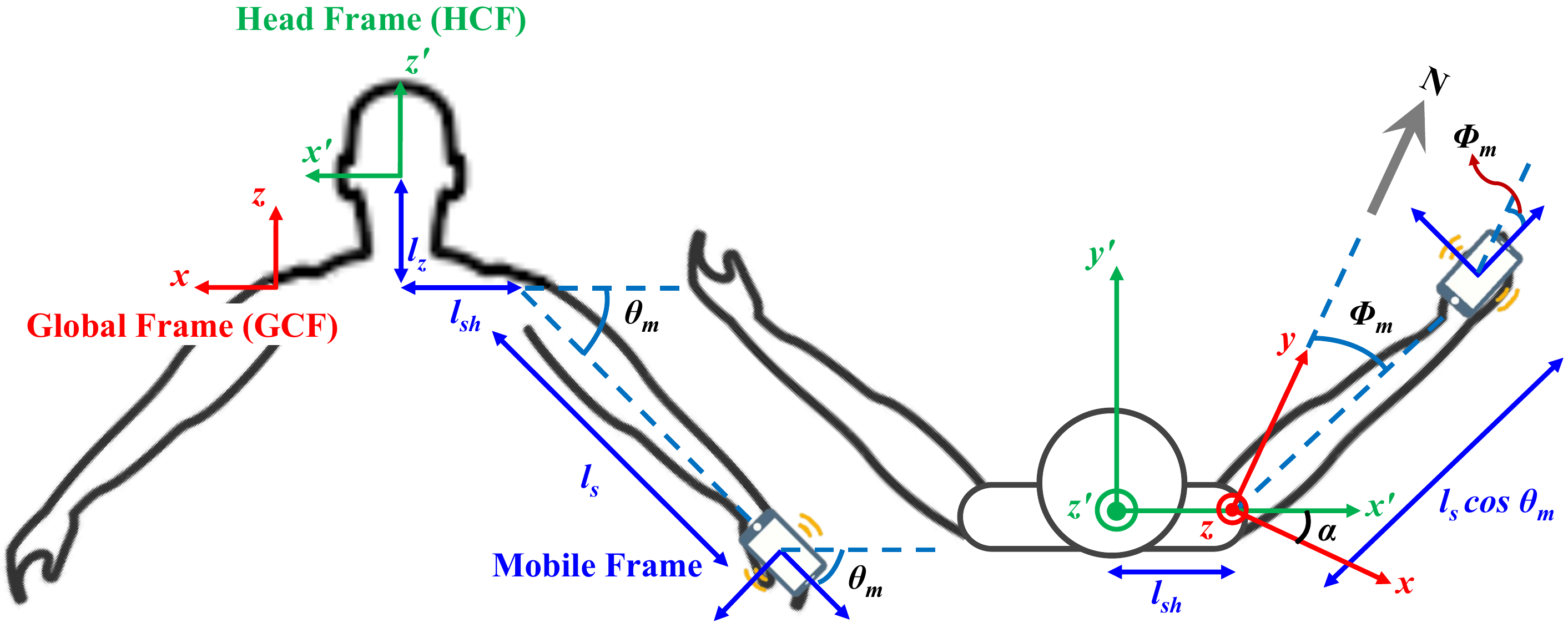}
    \caption{The notations used in the  sound direction finding algorithm.}
    \label{fig:body_}
    \vspace{-1em}
\end{figure}

\subsection{Estimating Sound Source Directions}
\input{direction_finding}

\subsection{Discussion}
A key design consideration of the measurement procedure is to ease user efforts and be tolerant to human errors. During measurements, deviations from the data collection protocol may impact the quality of data collected:
1) staying less than the required time in each location, and not capturing the full period of a chirp signal, 2) missing some required reference locations, and 3) performing measurements on one side of the body (e.g, using left or right hand only).

The system is by design robust to these deviations. First,  
since phone position are tracked automatically all the time, movements during reference sound playback can be detected. Data from locations with insufficient measurements are discarded. Second, using ITD and tracked phone orientations, we can determine whether measurements have been taken at all required reference locations. Otherwise, users will be alerted to move their phones to the missing locations. Third, in the same vein, the same procedure can tell whether measurements have been performed on both sides of a user’s body. If not, users will be promoted to conduct additional measurements.

\section{Performance Evaluation}
In this section, we evaluate the performance of the proposed HRTFs individualization procedure.

\begin{figure*}[!t]
     \centering
     \subfloat[$(-5^\circ,~24.48^\circ)$]{\includegraphics[width=1.8in]{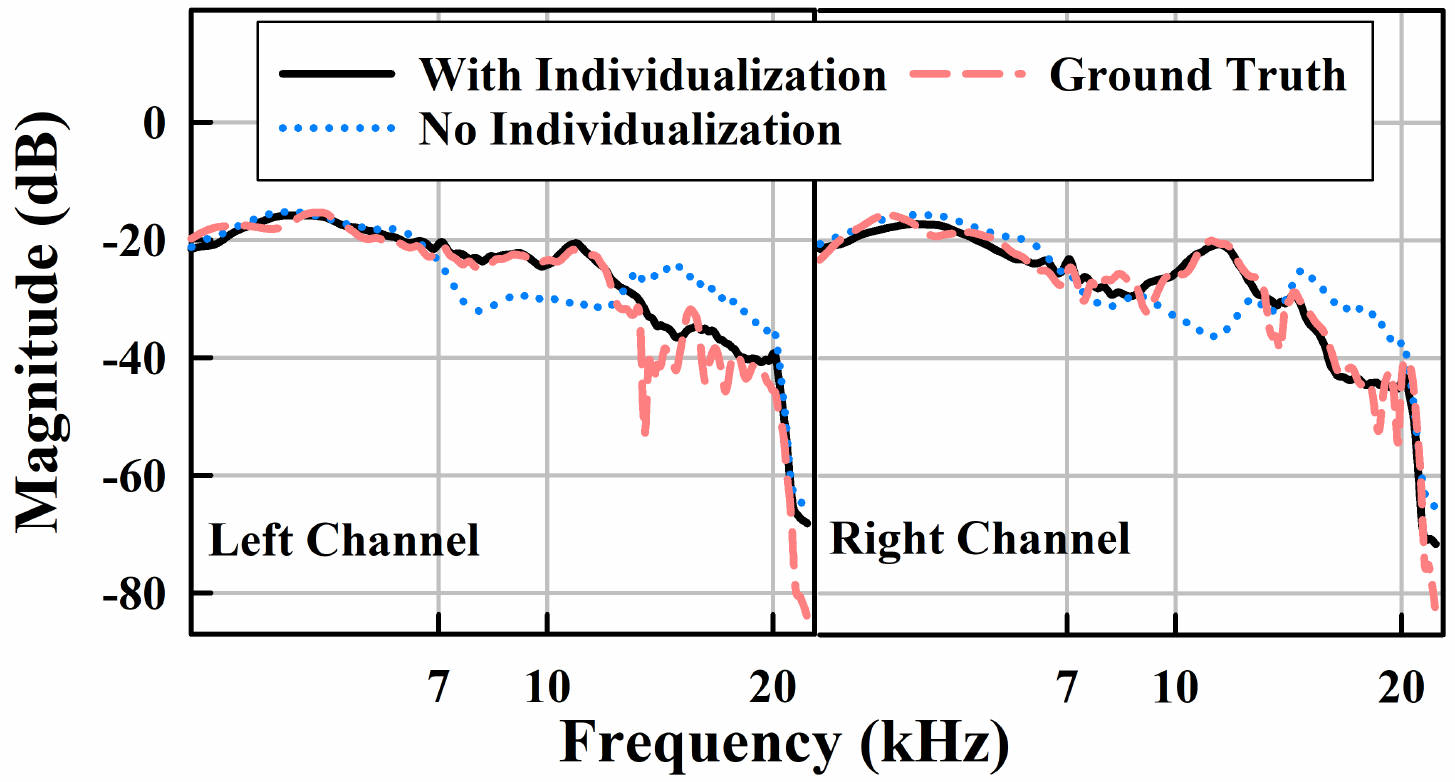}%
     \label{fig:pose_adapt_1}}
     \subfloat[$(50^\circ,~-30.96^\circ)$]{\includegraphics[width=1.8in]{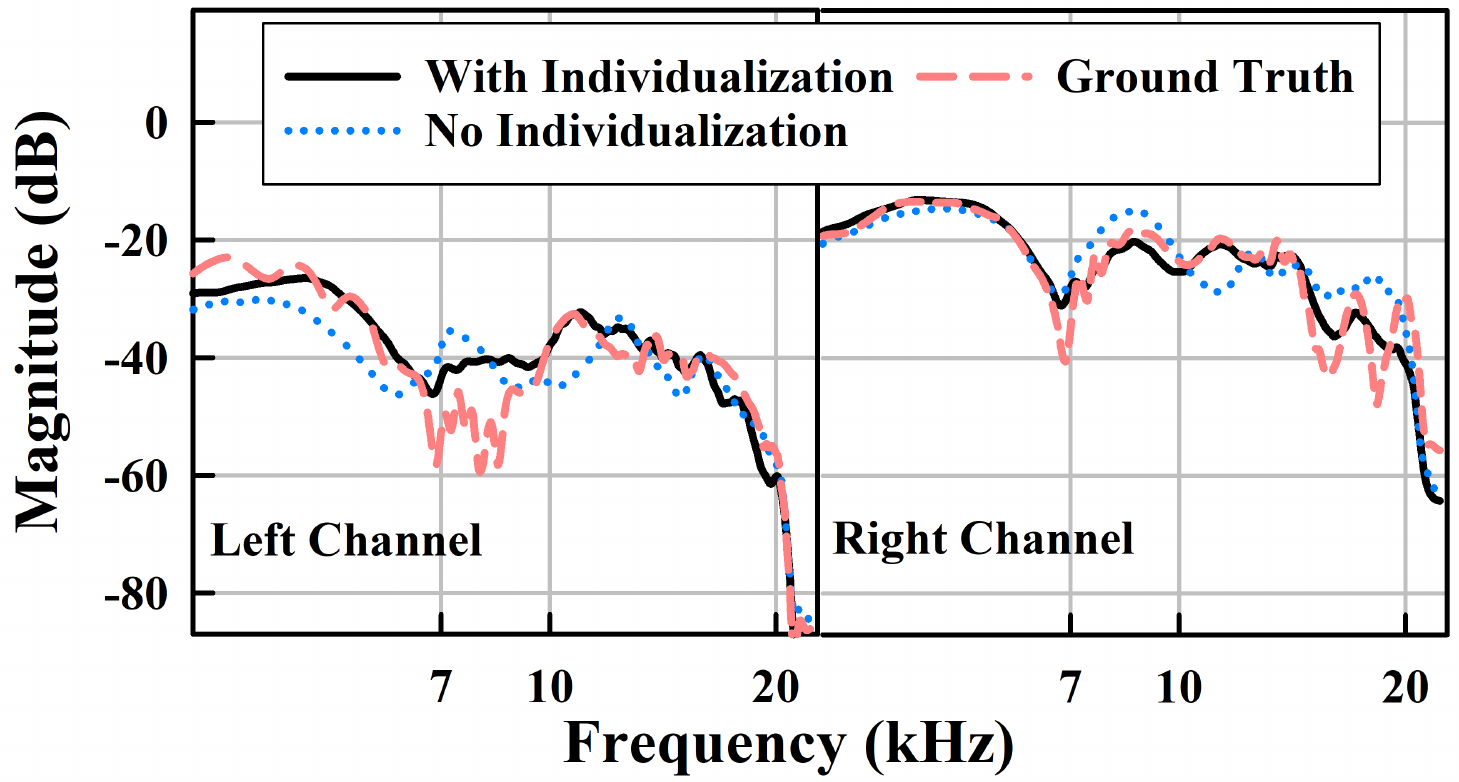}%
     \label{fig:pose_adapt_2}}
     \subfloat[$(-105^\circ,~14.40^\circ)$]{\includegraphics[width=1.8in]{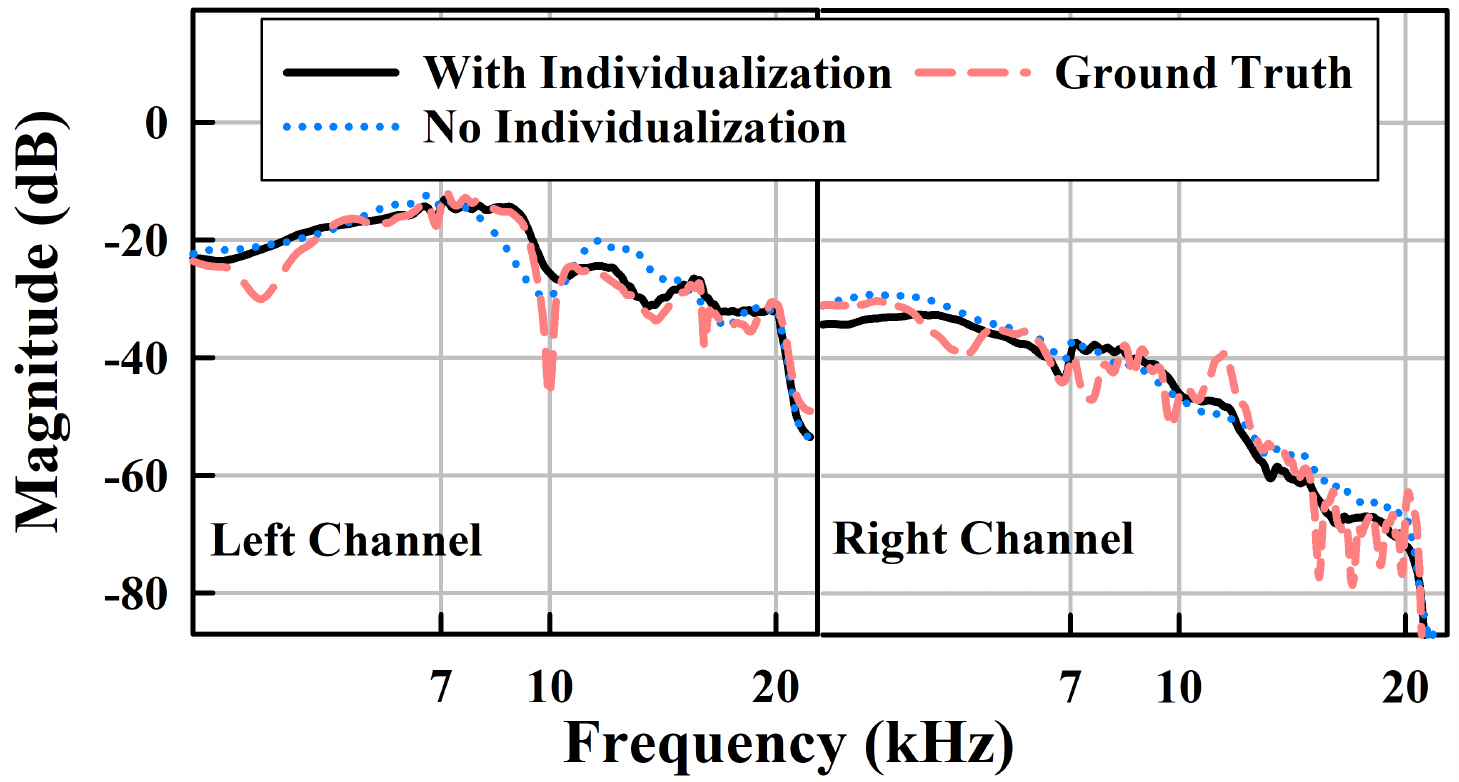}%
     \label{fig:pose_adapt_3}}
     \subfloat[$(-160^\circ,~-25.92^\circ)$]{\includegraphics[width=1.8in]{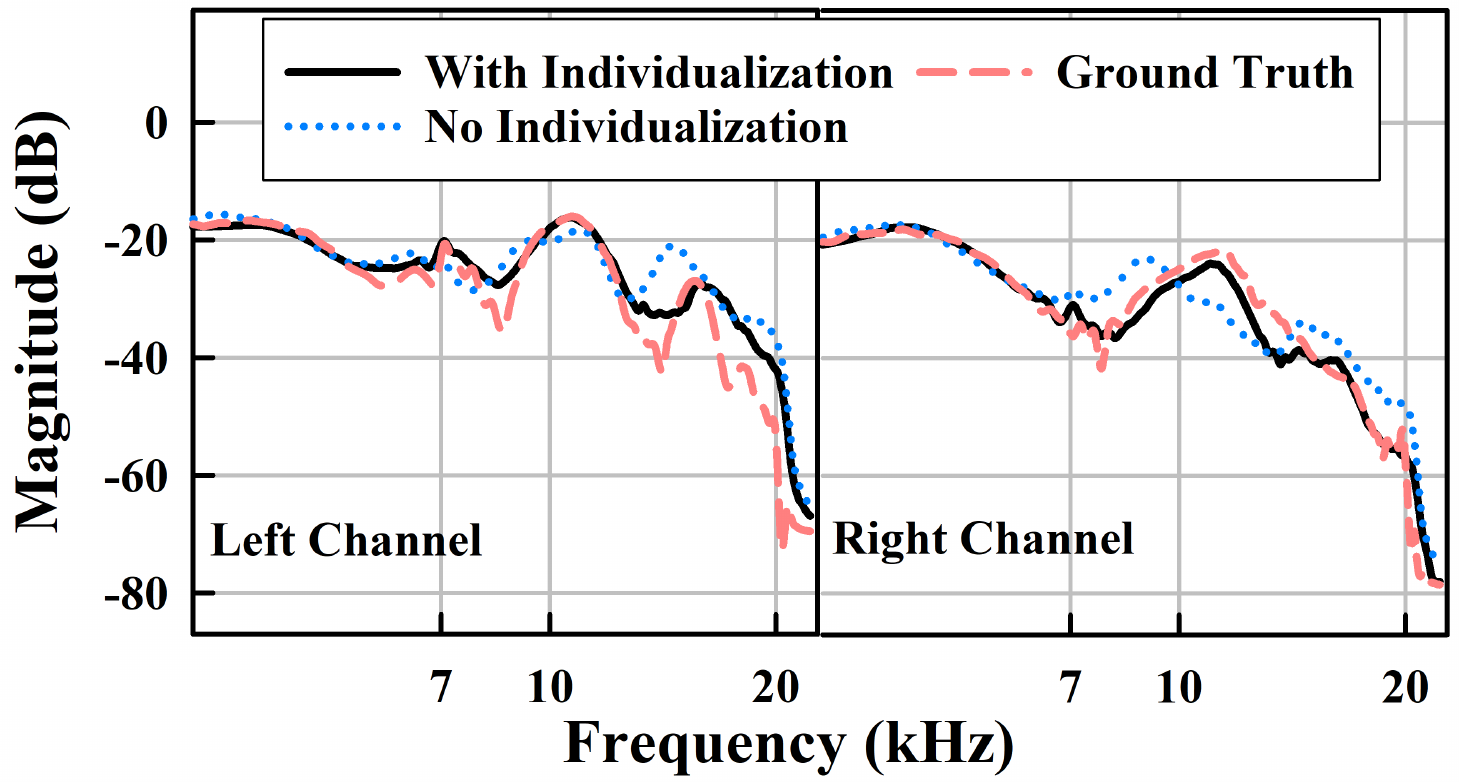}%
     \label{fig:pose_adapt_4}}
     \caption{Comparisons of ground truth HRTFs and HRTFs with and without individualization for Subject 1 from the ITA dataset at 4 different positions. Each curve concatenates the left and right HRTFs. The LSDs before individualization are: (a) 8.08, (b) 8.07, (c) 5.42, (d) 6.21, and after individualization (a) 4.62, (b) 4.25, (c) 3.47, (d) 4.14. Note, angles are *(Azimuth, Elevation).}
        \label{fig:CVAE_ADAPT}
\end{figure*}

\subsection{Implementation}
We implemented the proposed algorithm using Python and PyTorch. The system used for training and testing is equipped with Intel Xeon CPU @ 2.20GHz and 12GB RAM. Mini-batch Adam optimization with mini-batch size 128 and learning rate $10^{-4}$ are used during training. It takes approximately 3 hours on CPU for 100 iterations, and 11.4GB RAM to train the model on ITA dataset.
Decoder adaptation for a new subject consumes 1.6GB RAM, and takes 1 minute to run for 50 iterations.

\subsection{Validation on the ITA dataset}
During this step, we use the ITA dataset to evaluate the ability of the proposed CVAE model to generate HRTFs for subjects. Additionally, we study the effects of the number of measured directions and their spatial distribution on individualizing HRTFs for new users. Out of 48 subjects in the dataset, one subject is randomly chosen for adaptation and testing, while the remaining subjects are used for training. A small subset of the new user's data is used for adaption and the rest is used in testing. To quantify the accuracy of the predicted HRTFs, we utilize a metric called Log-Spectral Distortion (LSD) defined as follows \cite{rabiner1993fundamentals}: 
\begin{equation*} \label{eq_lsd}
\begin{split}
 LSD(H,\hat{H})=\sqrt{ \frac{1}{K} \sum_{k=1}^{K}{\bigg( 20 log_{10} \displaystyle\left\lvert \frac{H(k)}{\hat{H}(k)} \right\rvert \bigg)^2} }
\end{split}
\end{equation*}
where $H(k)$ and $\hat{H}(k)$ are the ground truth and estimated HRTFs in the frequency domain, respectively, and $K$ is the number of frequency bins. Clearly, if $H(k)$ and $\hat{H}(k)$ are identical, $LSD(H, \hat{H}) = 0$.

\begin{figure*}[!t]
\centering
\subfloat[LSDs for 11 subjects]{\includegraphics[width=2in]{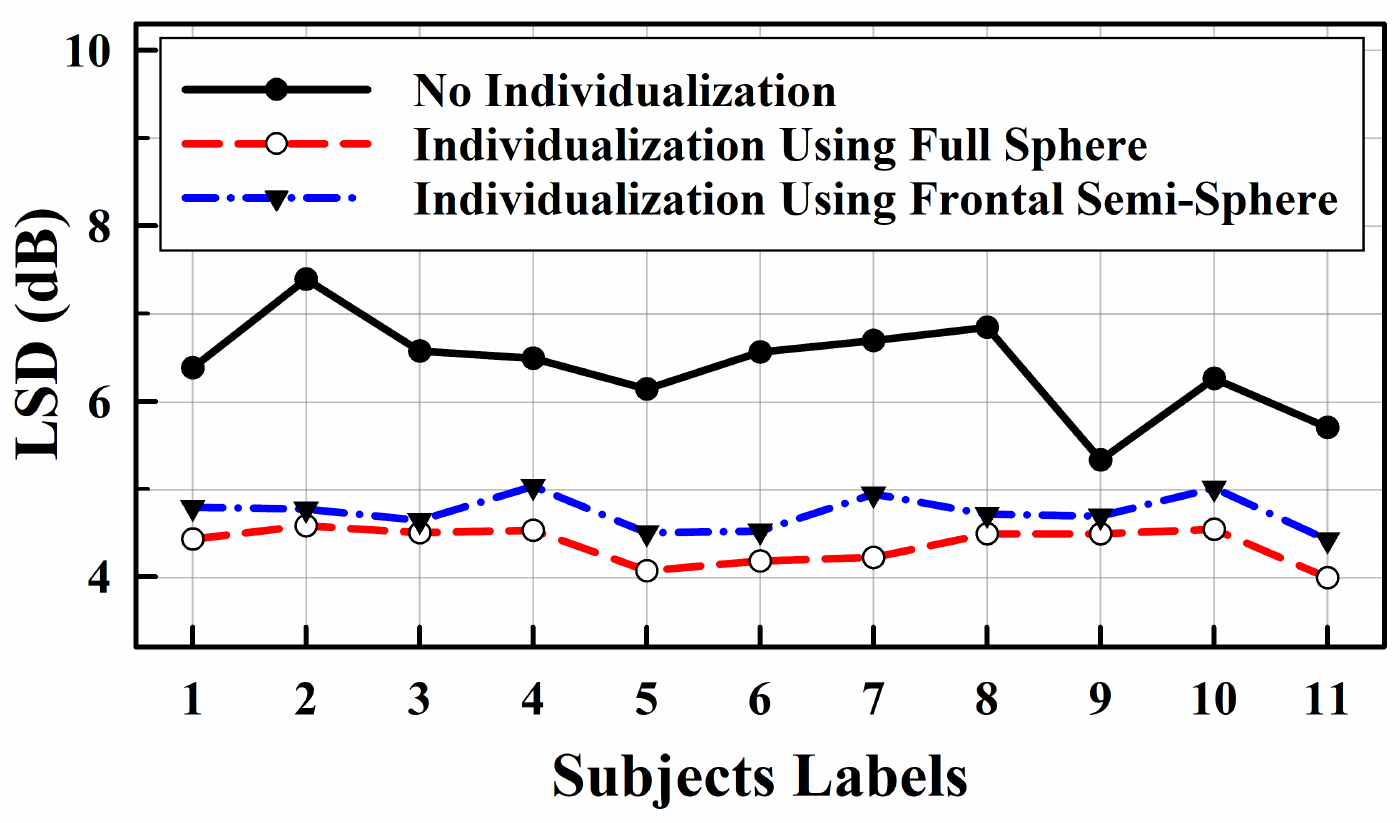}%
\label{fig:LSD_11subs}}
\hfil
\subfloat[Effects of measurements number]{\includegraphics[width=2in]{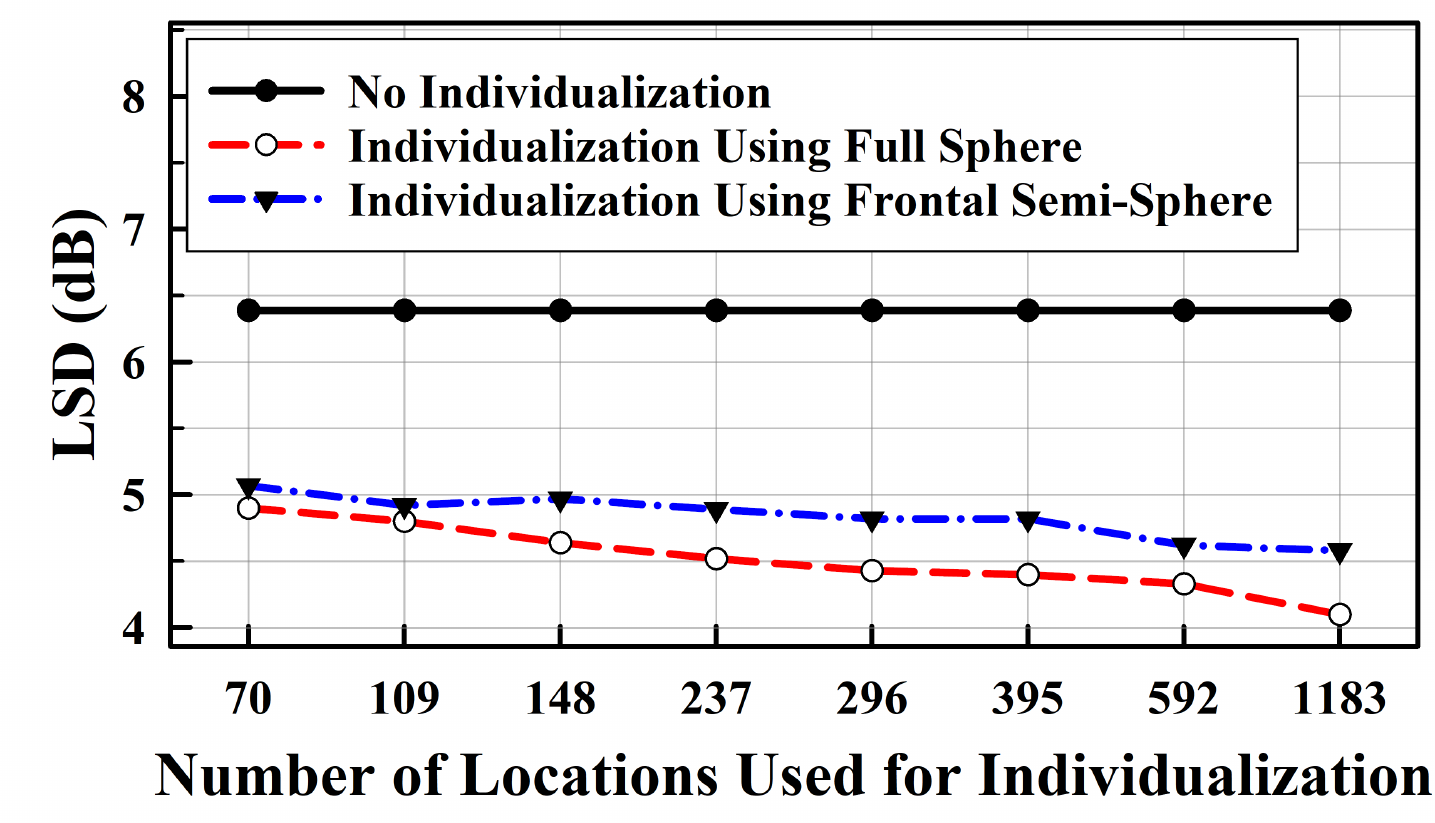}%
\label{fig:LSD_total}}
\hfil
\subfloat[Effects of the azimuth coverage range]{\includegraphics[width=2in]{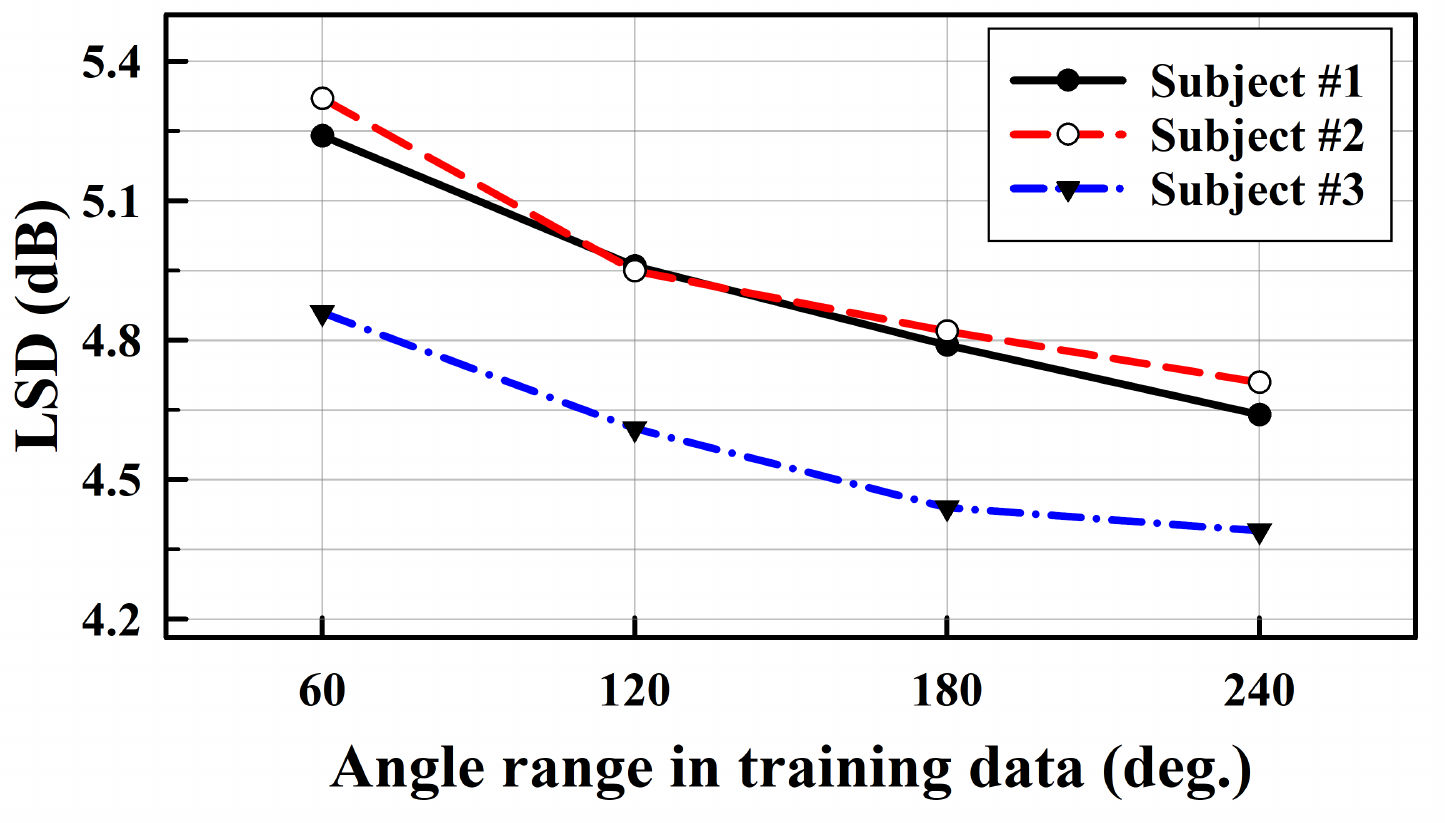}%
\label{fig:LSD_FoV}}
\caption{LSD errors for different subjects, from the ITA dataset, and with different measurement locations. In (a) and (b), individualization performance are shown in two cases: when the decoder is retrained using data only from the frontal semi-sphere (Blue) and using data from the full sphere (Red). (c) LSD errors for three subjects when the data used for individualization are chosen from a constrained azimuth angle range.}
\label{fig_10}
\end{figure*}

\emph{Fidelity of HRTF predictions}. Figure \ref{fig:CVAE_ADAPT} shows the generated HRTF for Subject 1 in the ITA dataset before and after adaptation in four different locations. In Figure \ref{fig:LSD_11subs}, the LSDs for 11 subjects in the ITA dataset are shown before and after adaptation. The lower LSDs after adaptation indicate that the proposed individualization model can successfully generate HRTFs for new users.

\begin{figure*}[!t]
     \centering
     \subfloat[$(-5^\circ,~24.48^\circ)$]{\includegraphics[width=1.8in]{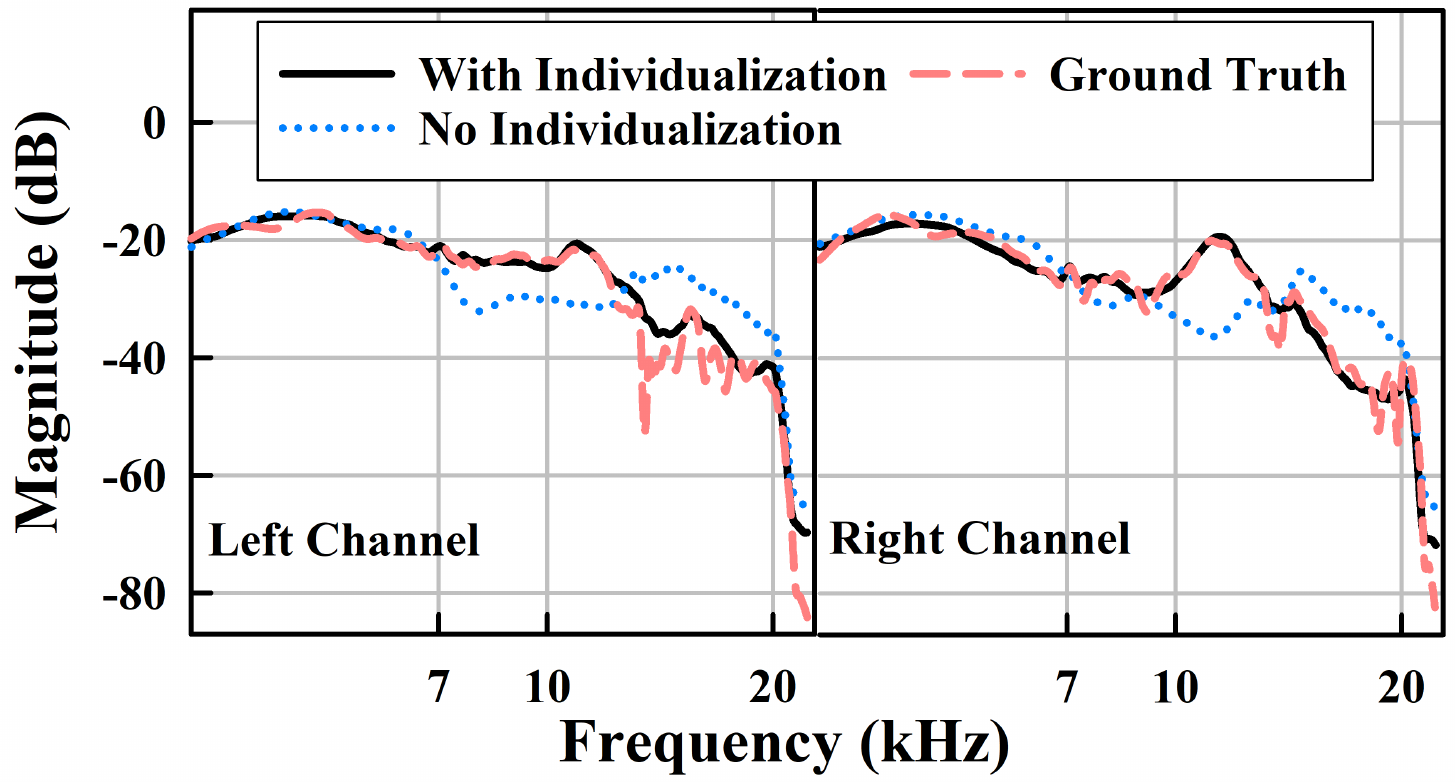}%
     \label{fig:body_adapt1}}
     \subfloat[$(50^\circ,~-30.96^\circ)$]{\includegraphics[width=1.8in]{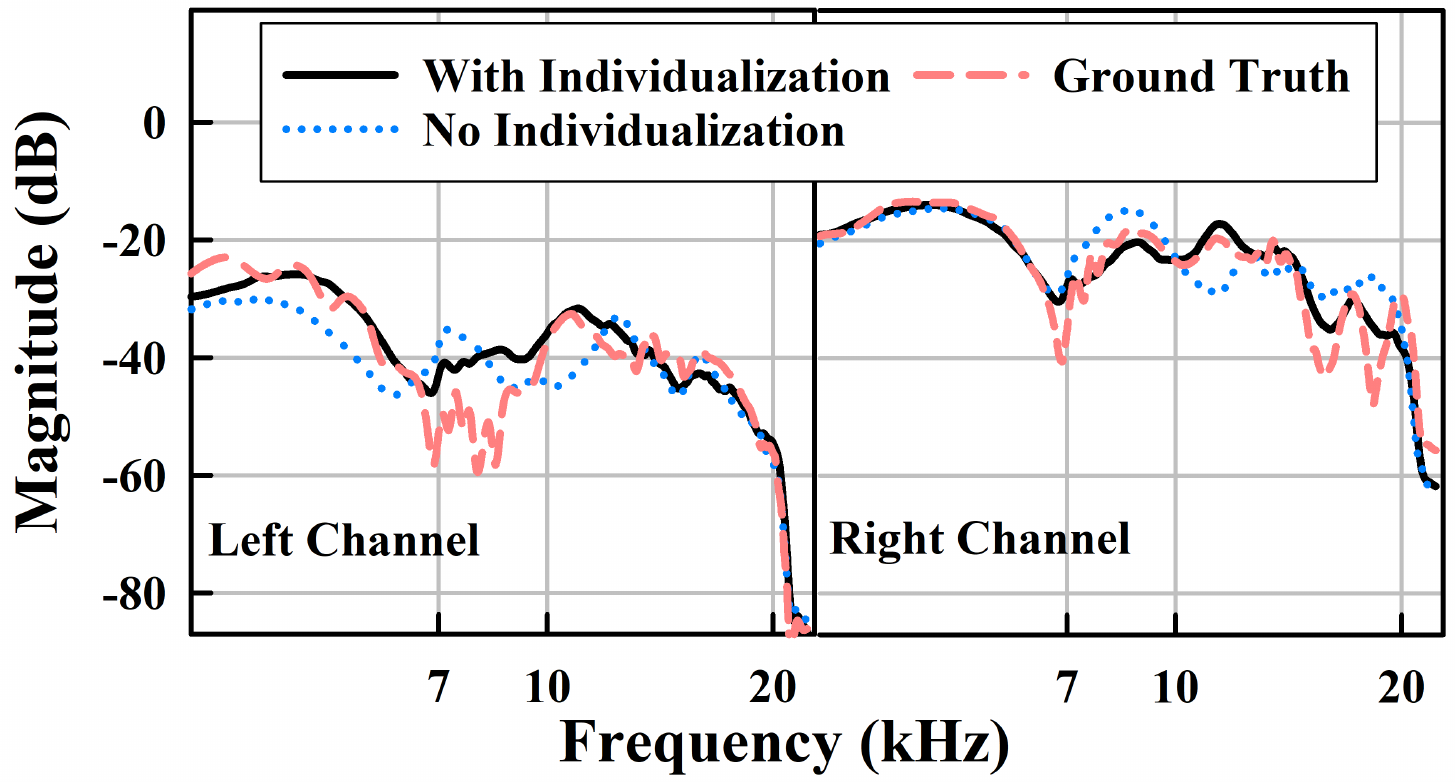}%
     \label{fig:body_adapt2}}
     \subfloat[$(-105^\circ,~14.40^\circ)$]{\includegraphics[width=1.8in]{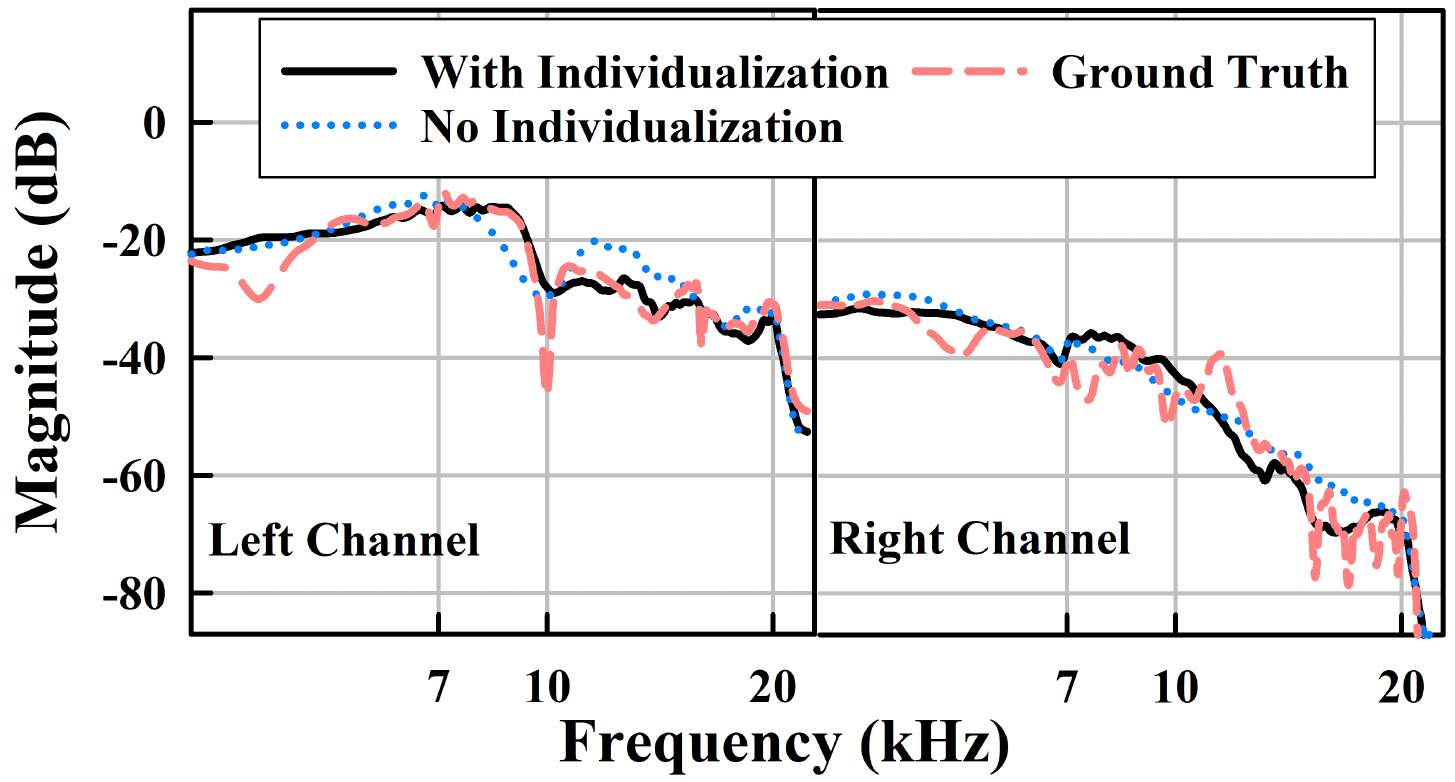}%
     \label{fig:body_adapt3}}
     \subfloat[$(-160^\circ,~-25.92^\circ)$]{\includegraphics[width=1.8in]{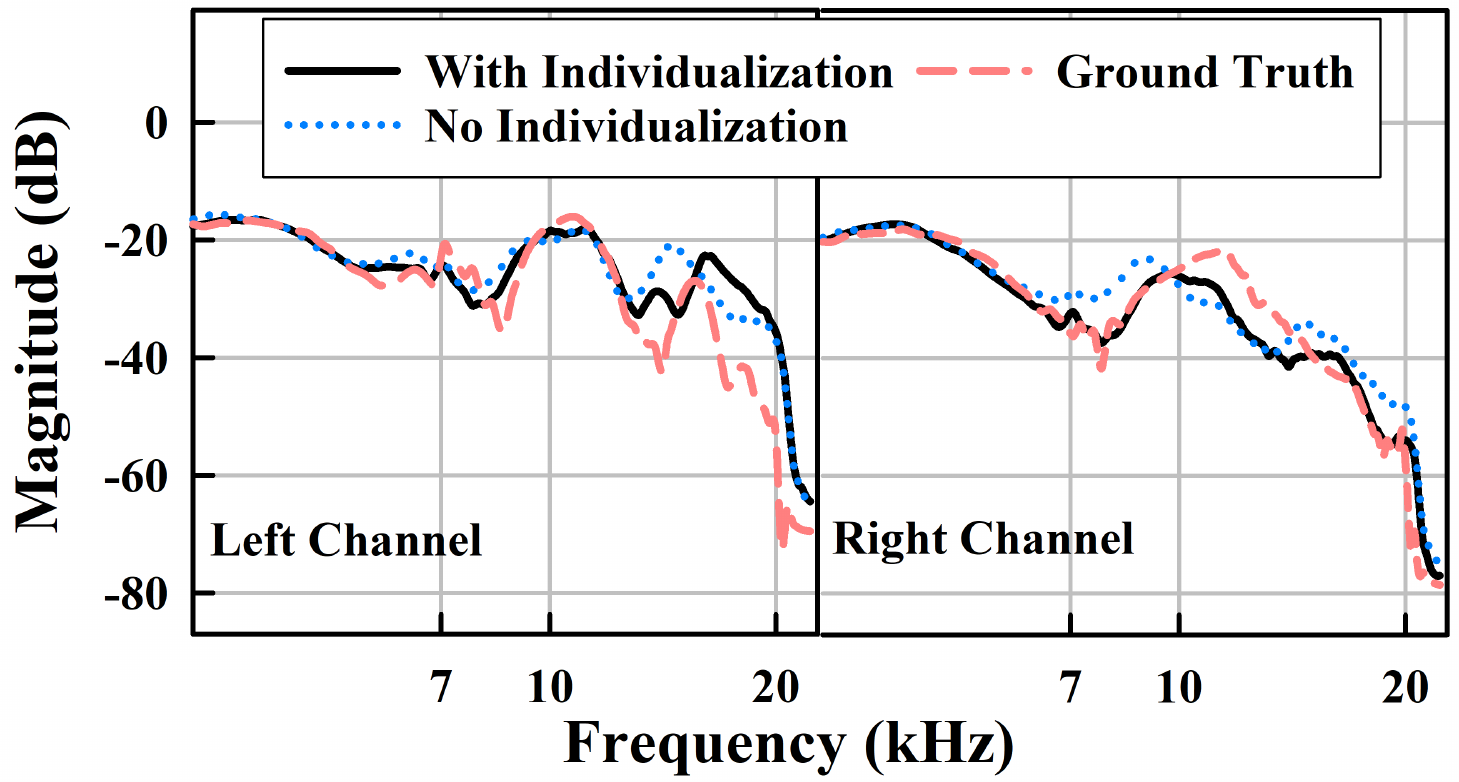}%
     \label{fig:body_adapt4}}
     \caption{Results of individualization using only HRTFs from locations in one's frontal semisphere. Each curve concatenates HRTFs from the left and right ears. The LSD errors before individualization are: (a) 4.62, (b) 6.64, (c) 7.41, (d) 7.37, and after individualization are: (a) 4.03, (b) 4.66, (c) 6.9, (d) 6.54. Note, angles are *(Azimuth, Elevation).}
        \label{fig:back_plane_result}
\end{figure*}

\emph{Effects of using measurements from frontal semi-spheres}. Recall in the procedure explained in Section \ref{real-data}, a subject moves her right and left hands holding a mobile phone to obtain sparse HRTF measurements. In absence of any measurement behind one's head, it is interesting to understand if our algorithm can fairly estimate HRTFs at back plane positions. To do this, we repeated the individualization step as before, but this time using the data from the frontal semi-sphere only. Figures \ref{fig:LSD_11subs} and \ref{fig:LSD_total} compare the individualization accuracy when data is chosen from the full sphere and the frontal semi-sphere. We observe that though LSDs increase compared to the case using the full-sphere data for individualization, significant improvement is still observed over non-individualization. Figure \ref{fig:back_plane_result} shows the estimated HRTFs with and without individualization, using data from the frontal semi-sphere. It is noticeable that the individualized HRTFS are more accurate than the case without individualization. However, we can observe less errors in the prdicted HRTFs in the frontal (Figure \ref{fig:body_adapt1}-\ref{fig:body_adapt2}) compared to those in the back (Figure \ref{fig:body_adapt3}-\ref{fig:body_adapt4}) semi-spheres.

\emph{Effects of the sparsity of measurement locations on individualization}. We have performed this test to define the minimum required number of measurements. As shown in Figure \ref{fig:LSD_total}, few measurement locations degrade the performance of individualization whether they are in the frontal semi-sphere or in the full sphere. However, with as little as 70-measurement locations, $20.7\%$ and $23.3\%$ reductions in LSDs can be achieved in the two cases, respectively.

\emph{Effects of the azimuth coverage on individualization}. We need to study these effects because the shoulder joints of different people may have different range of motion. Specifically, we take measurements only from locations whose azimuth angles fall in $[-\phi/2, +\phi/2]$, and vary $\phi$ from $60^\circ$ to $240^\circ$. Figure \ref{fig:LSD_FoV} shows the individualization results for three subjects. As expected, LSDs will drop when the azimuth coverage increases. However, even with measurements from only $\phi = 60^\circ$, LSDs are much less than those without individualization.

\begin{figure}[!t]
\centering
\subfloat[Train  Error]{\includegraphics[height=0.14\textwidth]{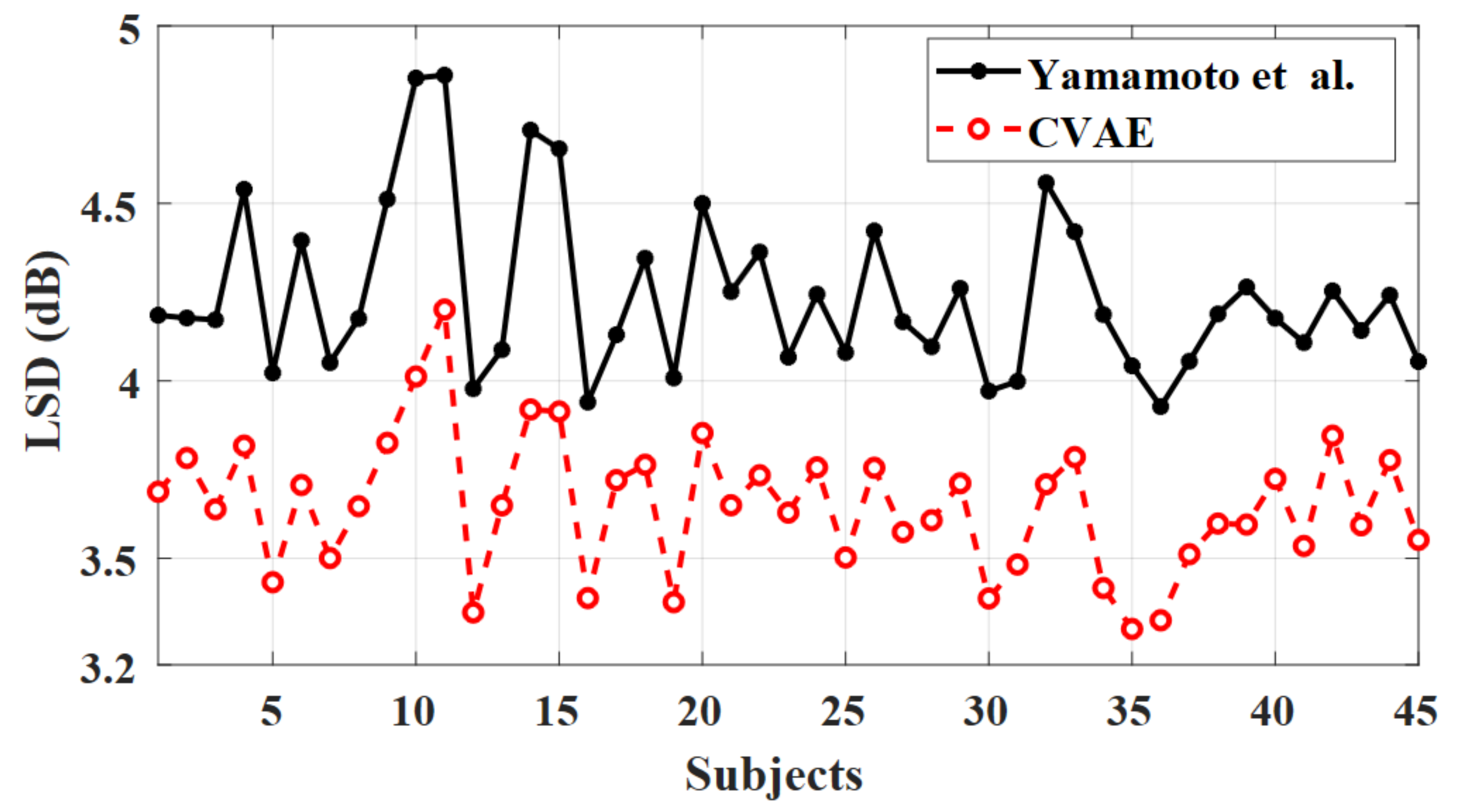}%
\label{ym_train}}
\hfil
\subfloat[Individualization Error]{\includegraphics[height=0.14\textwidth]{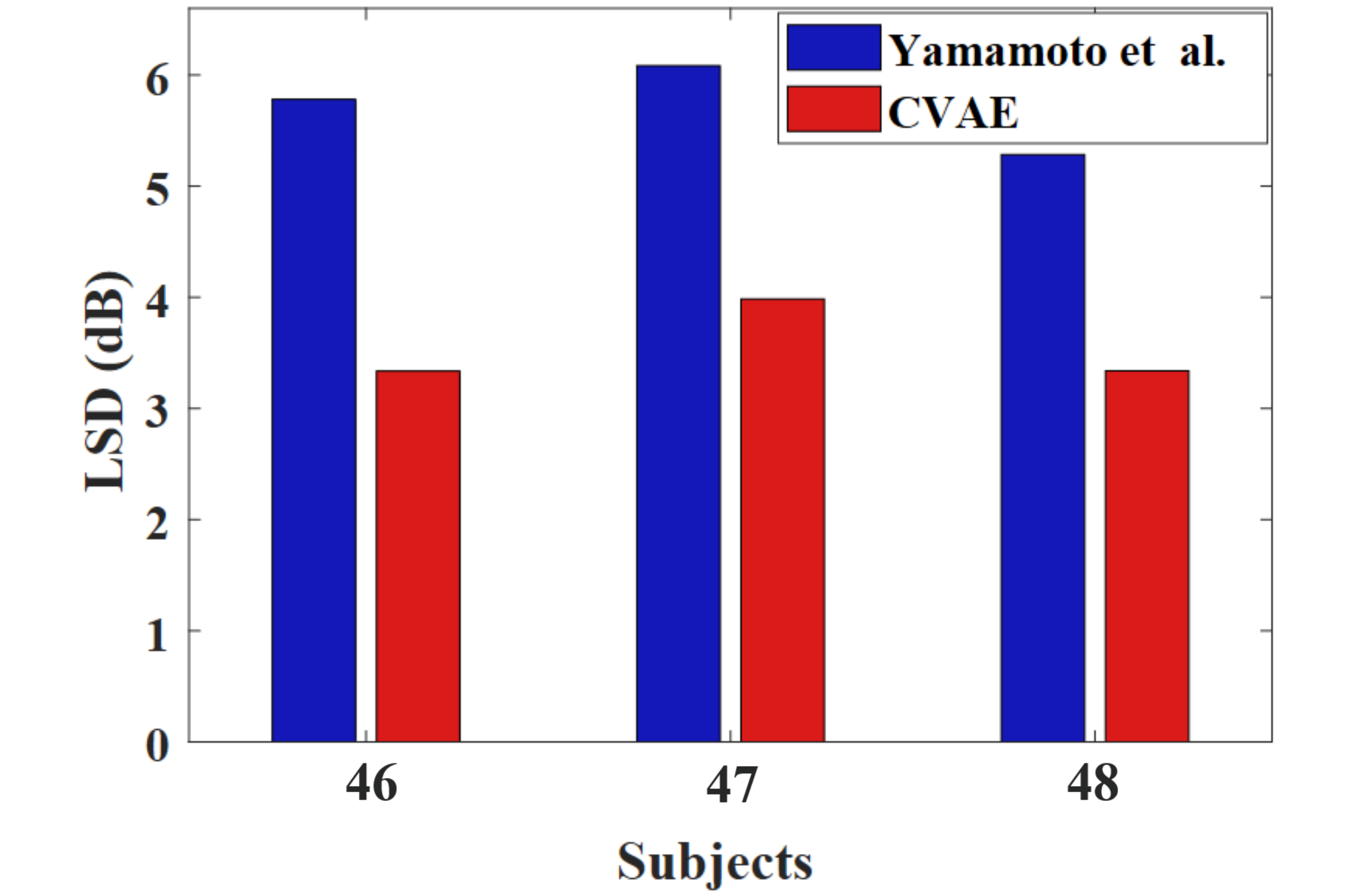}%
\label{ym_individ}}
\caption{Train  and individualization errors of the base model and the proposed model on different subjects. 45 subjects from ITA dataset are chosen. Both models are adapted for subjects 46 to 48 using data from all 2304 directions.}
\label{table:ym_error}
\vspace{-1em}
\end{figure}

\begin{figure}[!t]
\centering
\subfloat[$(-5^\circ,~24.48^\circ)$]{\includegraphics[width=0.23\textwidth]{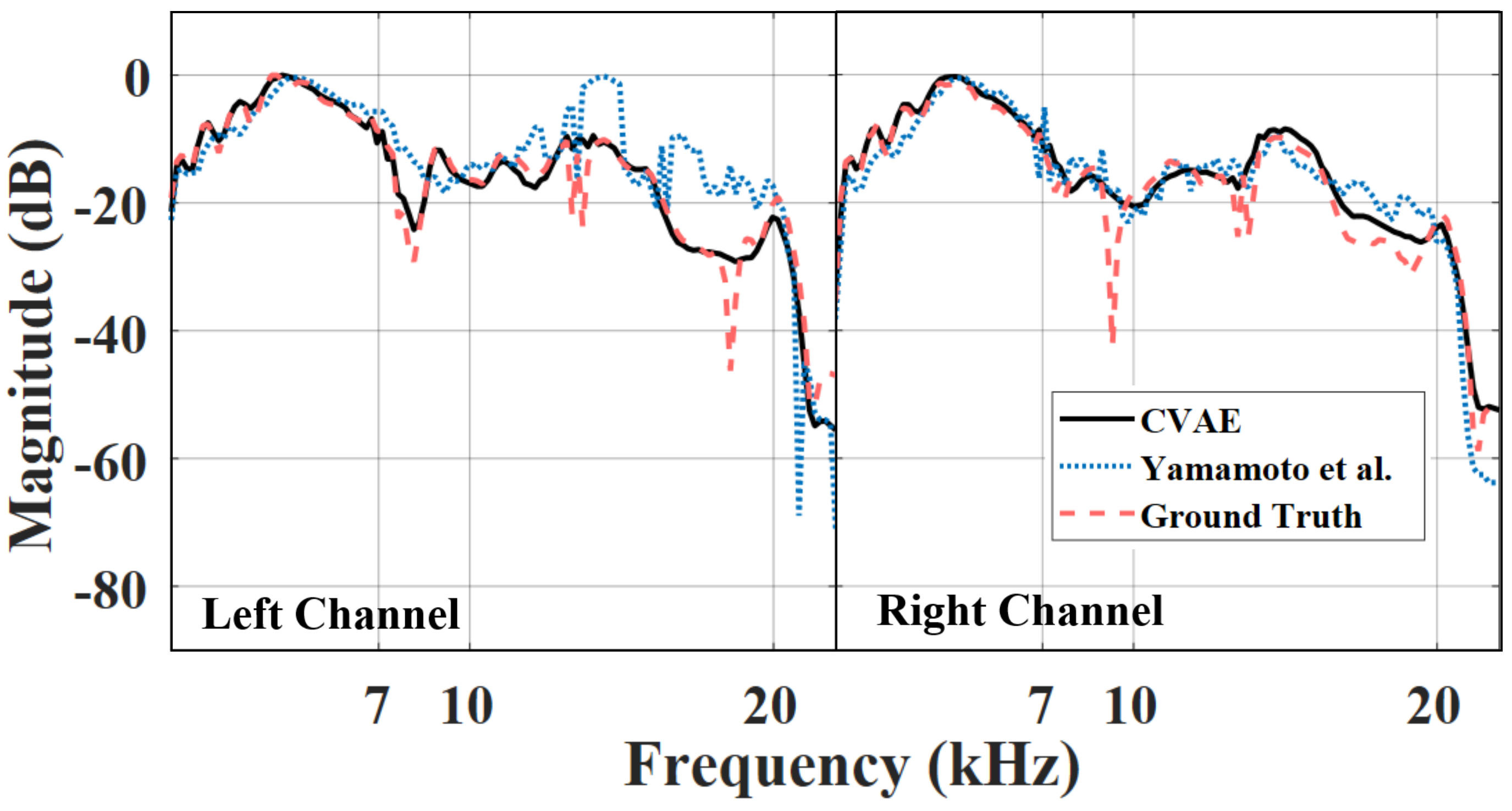}%
\label{tab_ym_cvae_a}}
\hfil
\subfloat[$(50^\circ,~-30.96^\circ)$]{\includegraphics[width=0.23\textwidth]{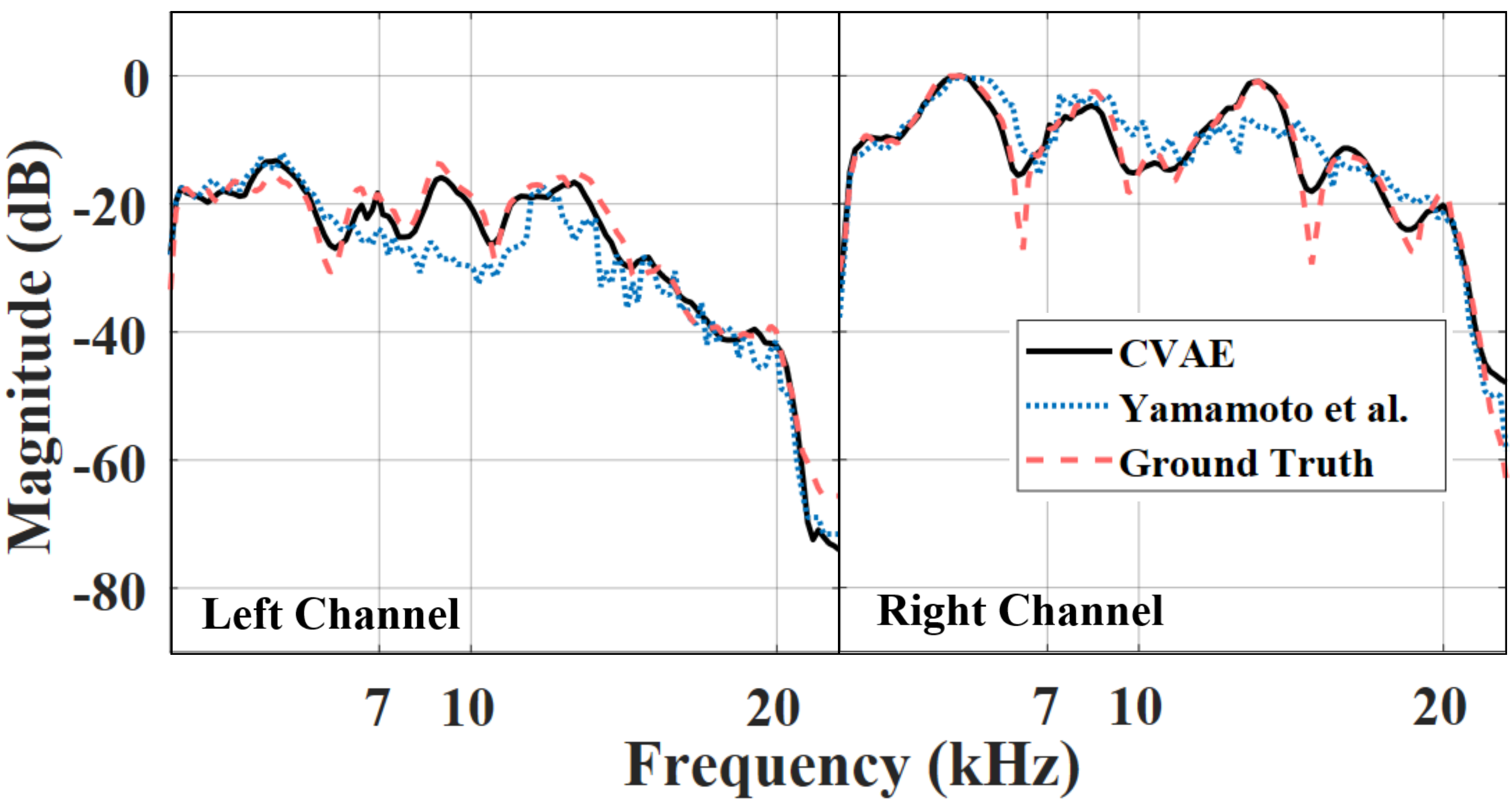}%
\label{tab_ym_cvae_b}}
\caption{
Result of individualization at two different (Azimuth, Elevation) directions for Subject 45 in the ITA dataset. Each curve concatenates the left and right HRTFs. The LSD errors for the proposed CVAE model are: (a) 3.22, (b) 2.79, and for Yamamoto's model are: (a) 7.01, (b) 5.39.}
\label{table:ym_cvae_comparison}
\vspace{-1em}
\end{figure}

\emph{Comparison with a state-of-the-art model}. We have implemented the algorithm in \cite{CVAE-02}, a state-of-the-art model for HRTF estimation. The architecture of the model in the work contains an Adaptive layer, which replaces ordinary Linear layers, and separates common factors and individuality of different subjects into two different tensors. We have followed the procedure described in the text in Appendix B in \cite{CVAE-02} to reproduce the architecture, and only compared the power spectrum output of their model with ours. To do so, we take the first 45 subjects in the ITA dataset, and train both the proposed CVAE model, and the model by Yamamoto et al. as the base model, on these subjects. Next, the models are individualized separately for each of the 3 remaining subjects, using data from all 2304 directions in the ITA dataset. To individualize the baseline model for new subjects, Covariance Matrix Adaptation Evolution Strategy (CMA-ES) optimization is applied to the subject vector, as mentioned in \cite{CVAE-02}. The results are shown in Figure \ref{table:ym_error}. It can be seen that both during training and in individualization, our proposed model outperforms the base model with smaller LSD. In Figure \ref{table:ym_cvae_comparison}, the result of individualization for subject 45 at two different directions are shown. Clearly, the proposed model results in HRTFs closer to the ground truth.

\subsection{Experimental Validation \label{experimental_valid}}
In the experiments, we evaluate the performance of the proposed algorithm with the real data using the setup described in Section \ref{real-data}\footnote{Research ethics approval has been obtained for user data collection.}. The purpose of the experiments is two-fold: 1) to evaluate the accuracy of the direction finding algorithm in Section \ref{real-data}, and 2) to evaluate the precision of the HRTF prediction model using real-world data.

\subsubsection{Implementation}
\begin{figure}[!t]
\centering
\subfloat[Mobile App]{\includegraphics[width=1.6in]{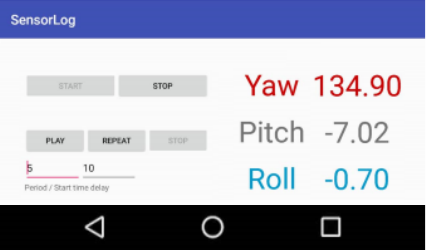}%
\label{fig:android_app}}
\hfil
\subfloat[In-ear electret microphones.]{\includegraphics[width=1.6in]{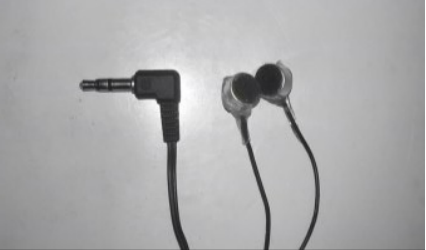}%
\label{fig:microphone}}
\caption{Software and hardware setup for data acquisition.}
\label{fig_12}
\end{figure}

\emph{Data Capture and Post-Processing}. An Android application has been developed (Figure \ref{fig:android_app}), with two main functions: 1) emitting reference sounds, and 2) logging the pose of the phone in its body frame (in yaw, roll, and pitch). The sweep time of the reference exponential sweep signal is 1.2 second, with frequencies range from 20Hz to 22kHz. With 1 extra second between consecutive measurements to let reverberations settle down, measuring 100 locations takes about 220 seconds, a little less than 4 minutes. Two electret microphones, soldered into a headphone audio jack (Figure \ref{fig:microphone}), are connected to a PC sound card for audio recording. The microphones are chosen to have good responses in human hearing ranges 20Hz$\sim$20kHz. Audacity \cite{audacity} is used on the PC for recording, which is an open-source audio software and has some preliminary editing features like noise reduction and normalization. We extract the HRIRs on MATLAB using the ITA Toolbox \cite{ITA-Toolbox_2017}. Note that with a suitable programming interface, the proposed system can be implemented on Bluetooth earphones that can stream recorded audio to a phone. Therefore, we can extract the HRIRs on the same sound emitting mobile phone.

\begin{figure}[!t]
    \centering
    \includegraphics[width=2.8in]{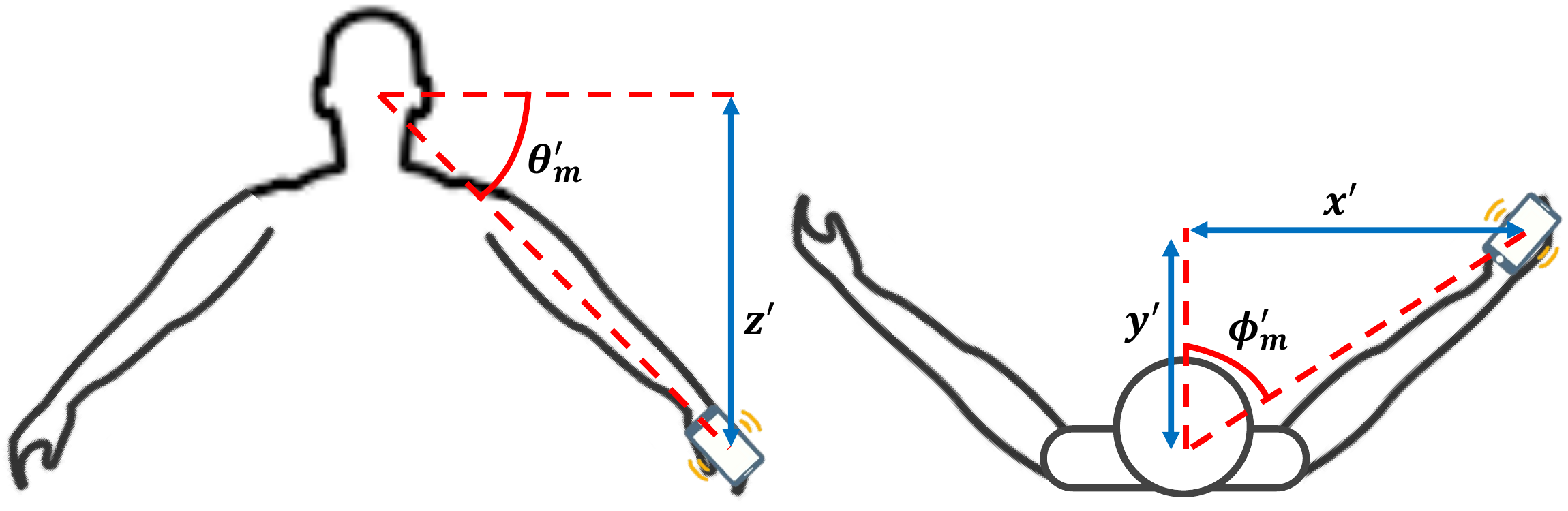}
    \caption{Ground Truth for directions. The measurement is done multiple times in different azimuths and elevations.}
    \label{fig:GT_labels}
\vspace{-1em}
\end{figure}

\emph{Ground Truth for Sound Source Directions\label{GT_LABEL}}. To evaluate the direction finding algorithm of sound sources, subjects are asked to stand on a marker on the ground, hold the mobile phone in their hand and point in different directions. At each position, we use a measurement tape to determine the vertical distance ($z'$) of the mobile phone to the centre of the subject's head, and its $x'$ and $y'$ coordinates in the horizontal plane, as shown in Figure \ref{fig:GT_labels}, with origin at the body center and $x$\emph{-axis} in the lateral direction and away from the body. Finally, we calculate the ground truth values of the azimuth and elevation angles as:
\begin{equation*} \label{eq_gt}
\phi_m' = \tan^{-1} \left(x' / y'\right) \quad \text{and} \quad \theta_m' = \tan^{-1} \left(z' / \sqrt{x'^2 + y'^2}\right)
\end{equation*}
The measurements are done for 10 different subjects, and one manikin, which is used to eliminate human errors such as undesired shoulder or elbow movements during measurements. The subjects were $5$ male and $5$ female with ages from $29$ to $70$, and heights from $158cm$ to $180cm$.

\begin{figure}[!t]
\centering
\subfloat[Azimuth angle estimation]{\includegraphics[width=0.23\textwidth]{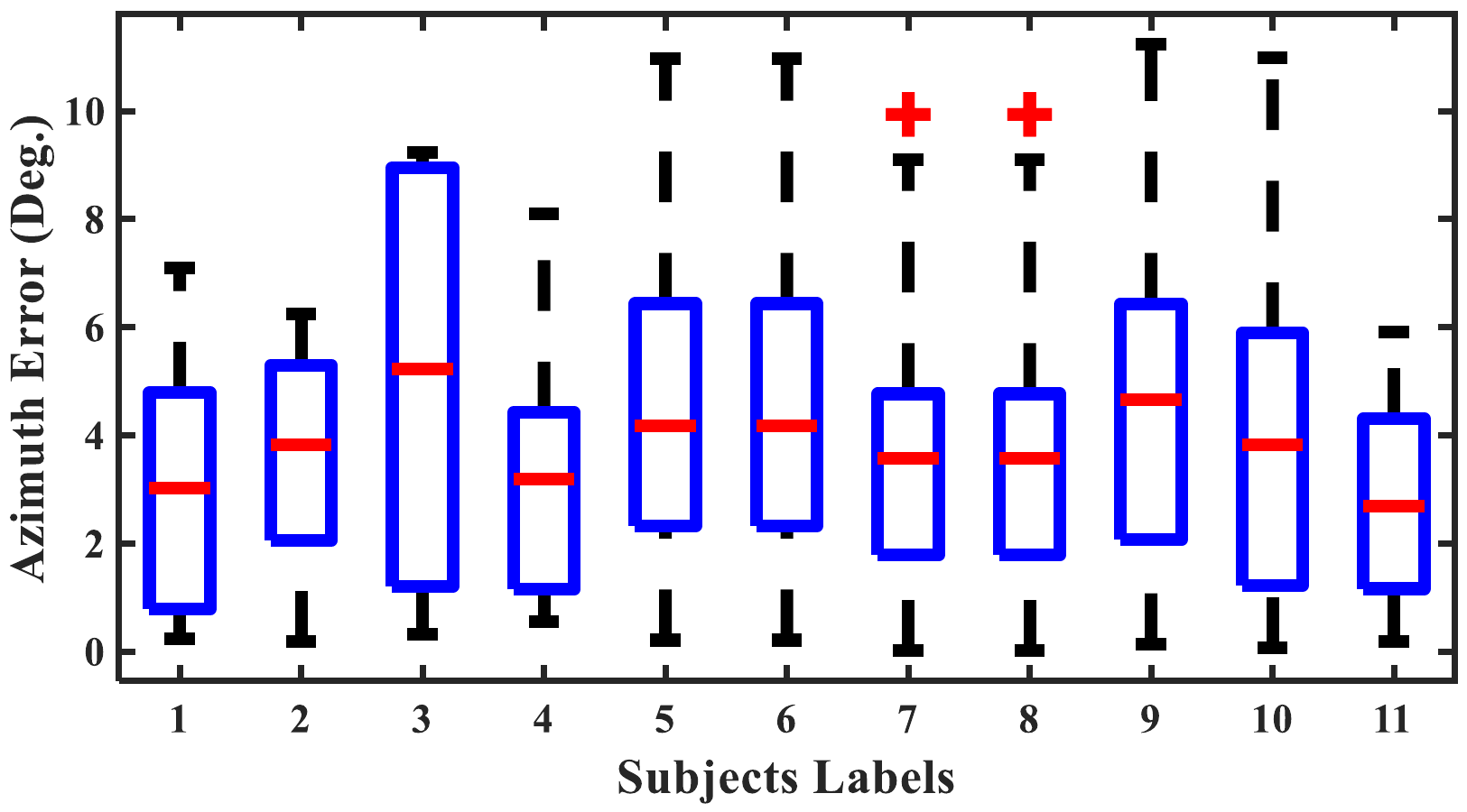}%
\label{tab_2a}}
\hfil
\subfloat[Elevation angle estimation]{\includegraphics[width=0.23\textwidth]{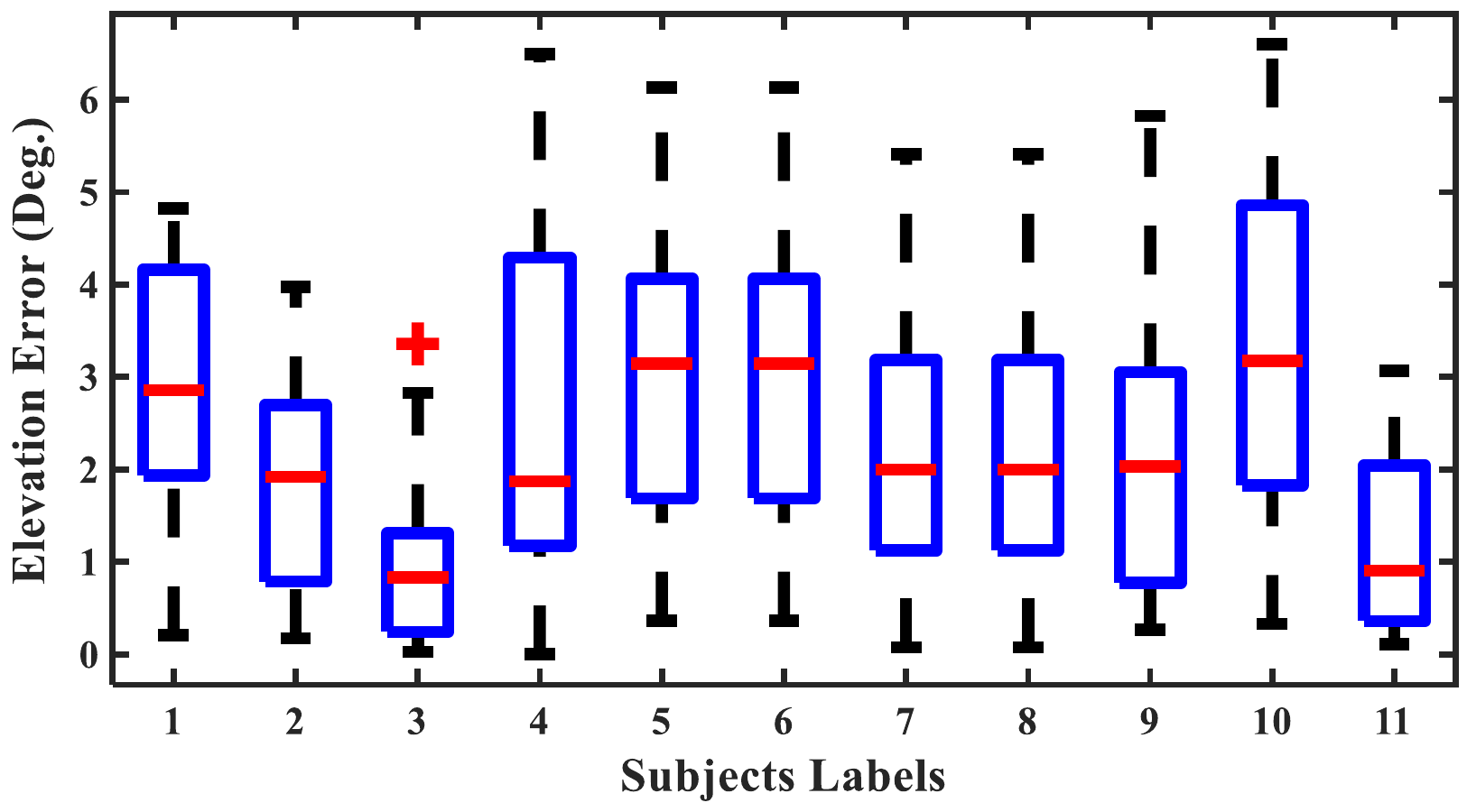}%
\label{tab_b}}
\caption{Direction finding estimations for different subjects. Labels from 1 to 10 are for the human subjects, while Label 11 is for the manikin. For each box, the red line is the median, and the bottom and top edges indicate the 25th and 75th percentiles, respectively.}
\label{table:1}
\vspace{-1em}
\end{figure}

\subsubsection{Results for direction finding\label{direction_find}}
Figure \ref{table:1} show the median, 25th and 75th percentiles of azimuth and elevation angles estimations using the algorithm in Section \ref{real-data}. Generally, we observe larger errors in azimuth than in elevation. This may be attributed to a larger range of motions horizontally (with both hands). By eliminating shoulder and elbow movements, the use of a manikin  leads to the least angle estimation errors as expected, demonstrating the correctness of the proposed algorithm. More detailed results for one subject for estimations at different sound source locations are given in Table \ref{table:2}. Note even when the phone is at the same height, due to distance between the subject's shoulder joint and head center, the elevation angles can differ.

\subsubsection{HRTF predictions}
\begin{figure*}[!t]
     \centering
     \subfloat[$(69.2^\circ,~10.92^\circ)$]{\includegraphics[width=1.8in]{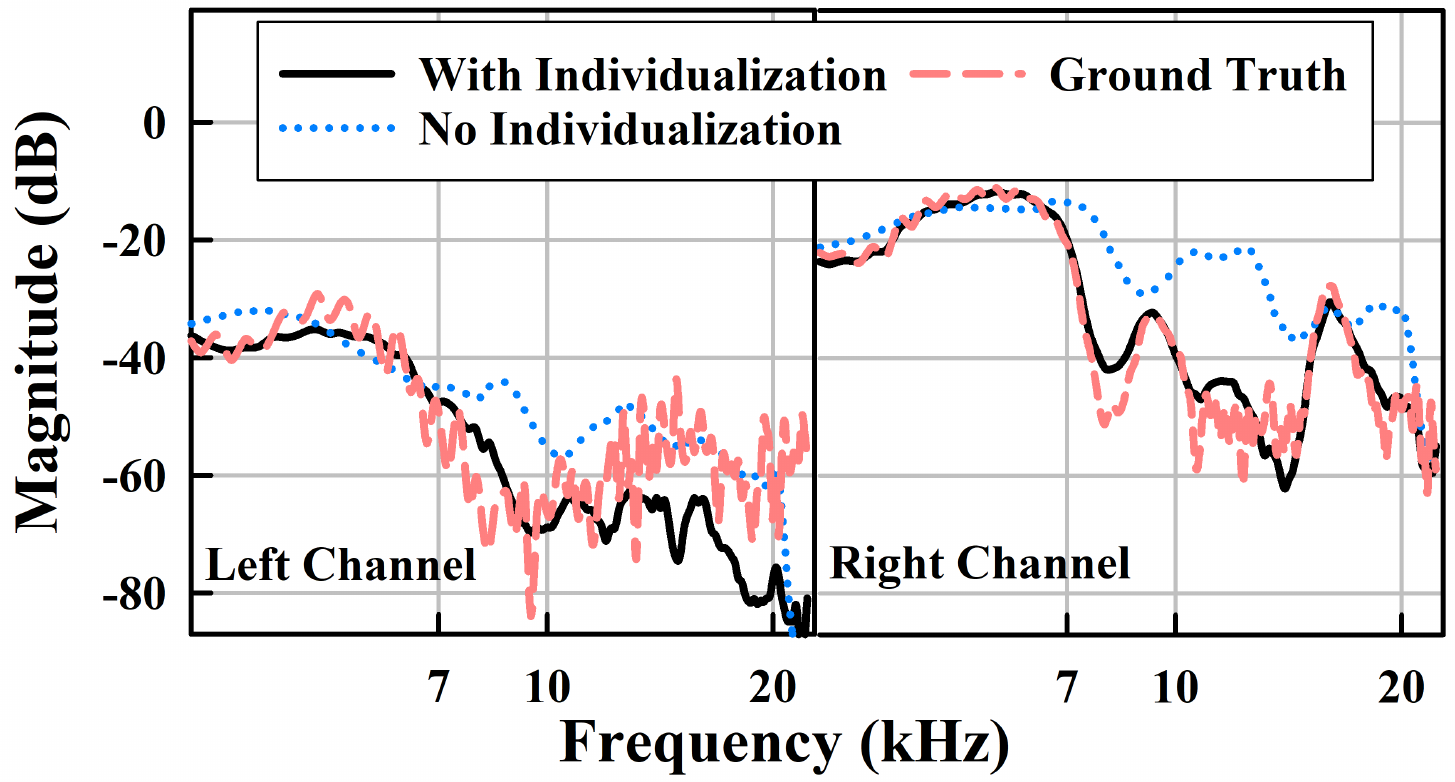}%
     \label{fig:measured_Adapt1}}
     \subfloat[$(5.71^\circ,~10.84^\circ)$]{\includegraphics[width=1.8in]{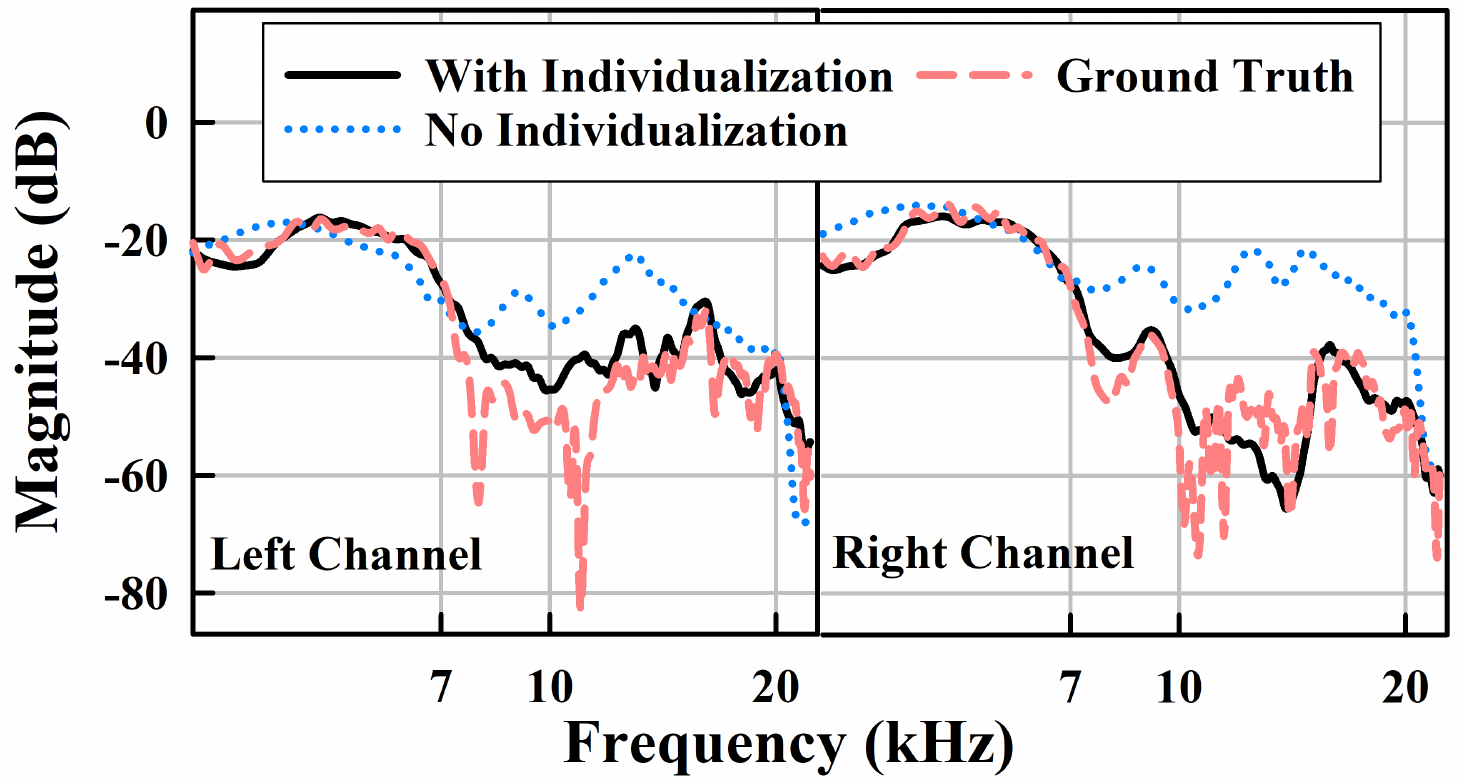}%
     \label{fig:measured_Adapt2}}
     \subfloat[$(41.69^\circ,~0.38^\circ)$]{\includegraphics[width=1.8in]{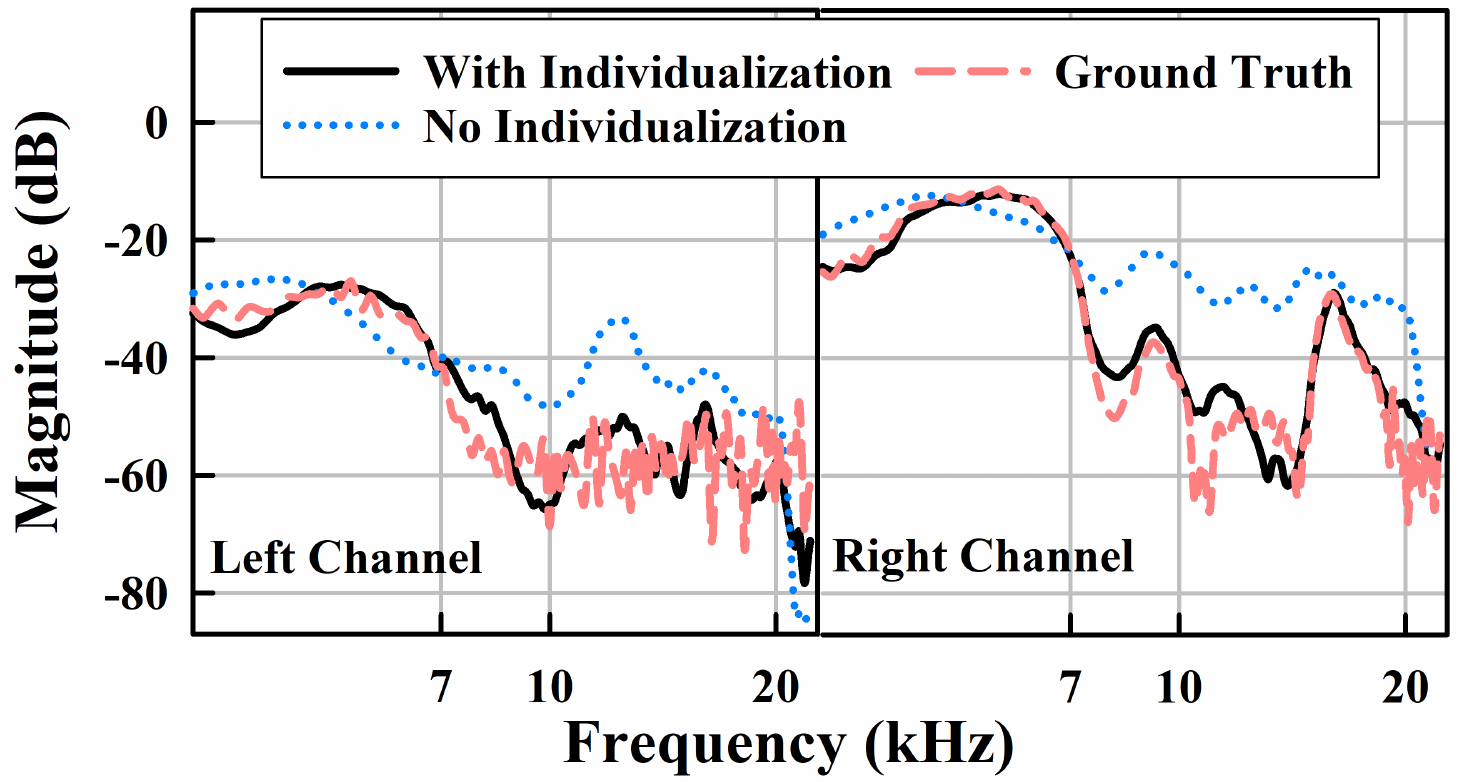}%
     \label{fig:measured_Adapt3}}
     \subfloat[$(12.7^\circ,~-2.71^\circ)$]{\includegraphics[width=1.8in]{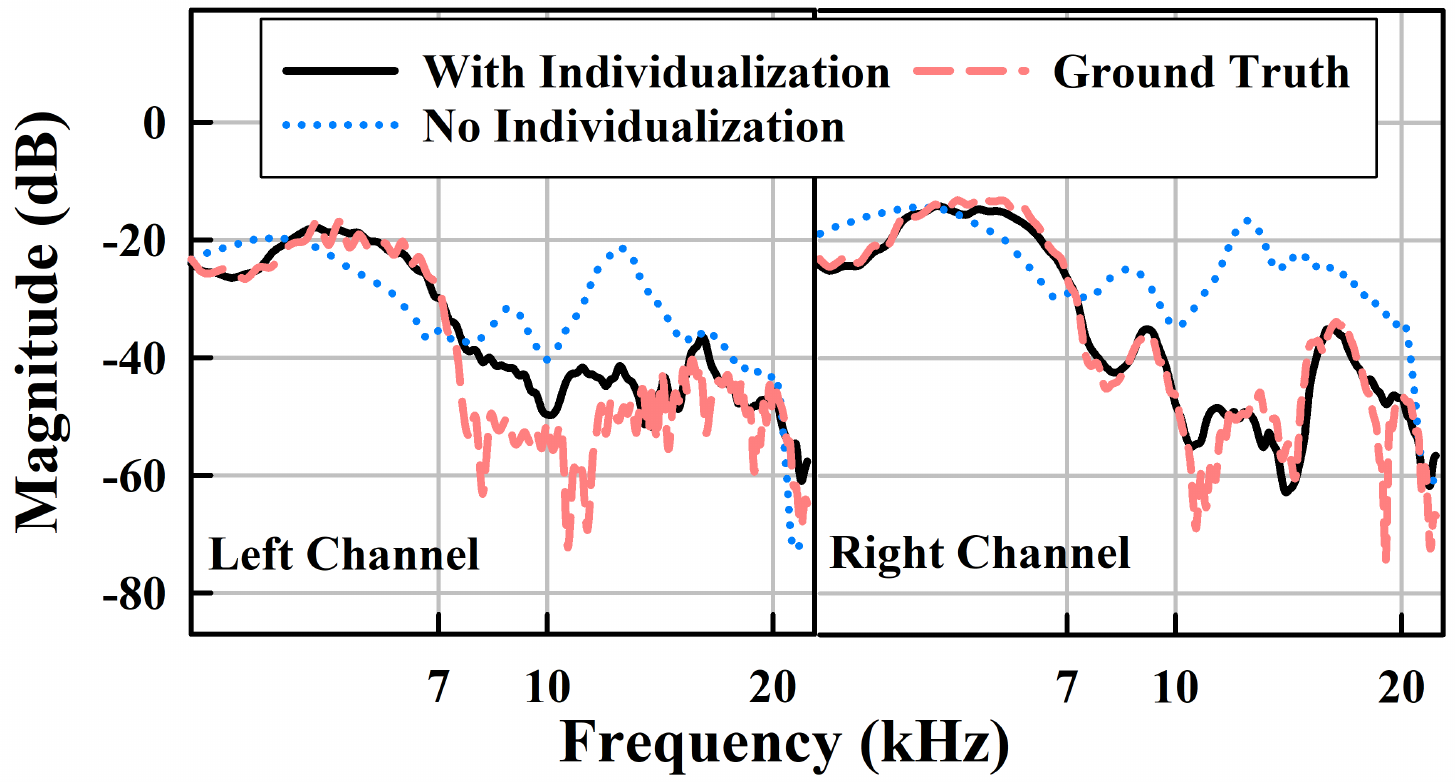}%
     \label{fig:measured_Adapt4}}
     \caption{Individualization using measurements data from one subject. 60 out of 83 measured locations are used for individualization. Each curve concatenates HRTFs from the left and right ears. The LSD errors before individualization are: (a) 13.79, (b) 15.48, (c) 15.03, (d) 16.10, and after individualization are  (a) 7.61, (b) 7, (c) 6.53, (d) 7.07. Note, angles are *(Azimuth, Elevation).}
        \label{fig:measured_data}
\vspace{-1em}
\end{figure*}

The results of individualization for one test subject are shown in Figure \ref{fig:measured_data}. For this subject, measurements at 83 locations were collected during the experiment, 60 of which were used for individualization, and the remaining 23 locations were used for testing. The individualized HRTFs clearly resemble the measured one more closely than without individualization in all cases. Moreover, we have performed the measurements in two different places, one of the team members home and our lab. For the home, we have taken the measurements for four participants, 2 females and 2 males, and the rest of the measurements have been taken in our lab. The used microphones, mobile phone and PC in the home were different than the used ones in our lab. We observe that the accuracy of the measurements, for HRIRs and directions, are the same from these different setup conditions.

\begin{table}[!t]
\centering
\caption{Estimated azimuth and elevation angles of sound sources  for one subject. 'Height' is the vertical difference between the subject's head center and the sound source.}
\begin{center}
\begin{tabular}{|c||c|c||c|c|}
 \hline
 Height & Azimuth & Error & Elevation & Error\\
 \hline 
  \multirow{4}{*}{$50.5cm$} & $26.98^{\circ}$ & $-3.49^{\circ}$ & $48.71^{\circ}$  & $-2.17^{\circ}$ \\
  & $49.81^{\circ}$ & $-5.64^{\circ}$ & $51.64^{\circ}$  & $-1.00^{\circ}$ \\
  & $73.87^{\circ}$ & $-1.86^{\circ}$ & $54.29^{\circ}$  & $-3.60^{\circ}$\\
  & $100.48^{\circ}$ & $0.18^{\circ}$ & $57.9^{\circ}$ & $-2.14^{\circ}$ \\ 
  \hline
  \multirow{4}{*}{$33.0cm$} & $4.04^{\circ}$ & $-5.42^{\circ}$ & $30.64^{\circ}$  & $-0.44^{\circ}$ \\
  & $55.02^{\circ}$ & $-0.94^{\circ}$ & $33.46^{\circ}$ & $-1.92^{\circ}$ \\
  & $80.47^{\circ}$ & $3.63^{\circ}$ & $36.65^{\circ}$ & $-3.19^{\circ}$ \\
  & $111.49^{\circ}$ & $6.24^{\circ}$ & $39.99^{\circ}$ & $-3.97^{\circ}$ \\
  \hline
   \multirow{3}{*}{$21.0cm$} & $2.98^{\circ}$ & $-5.98^{\circ}$ & $19.63^{\circ}$  & $-0.36^{\circ}$ \\
   & $51.23^{\circ}$ & $-1.89^{\circ}$ & $20.95^{\circ}$ & $-1.83^{\circ}$ \\
   & $101.18^{\circ}$ & $-4.66^{\circ}$ & $25.14^{\circ}$ & $-3.49^{\circ}$ \\
 \hline
\end{tabular}
\end{center}
\label{table:2}
\vspace{-1em}
\end{table}

It is worth mentioning that since measurements are done in an indoor environment, the calculated HRTF is a combination of room effects, HRTFs of the test subjects and distortions of the speaker and the microphones. Despite so, the results are still meaningful since applications of HRTFs such as binaural localization need to account for environment effects. Since the data acquisition procedure for individualization is fast and simple, one can reasonably do so for each new environment. As part of our future work, we will develop procedures to remove distortions from measurement devices and room impulse responses.

\section{Applications of Individualized HRTFs}
In this section, we implement two applications and demonstrate the advantage of HRTFs individualization. 

\subsection{Binaural localization}
Many different algorithms have been developed for binaural sound localization including simple regression models like SVR \cite{SVR-01}, GPR \cite{GPR-02}, and deep neural network models \cite{DL-01}. In this paper, we follow a state-of-the art neural network-based model in \cite{DL-01}, to evaluate the benefits of accurate HRTF predictions. We believe the same observations are applicable to other localization algorithms.

\emph{Localization model}. The model is a fully-connected neural network with four hidden layers. Each hidden layer has 500 units, with a ReLU activation function, and a dropout layer after. The output is a classification over 72 discrete azimuth angles. The network takes as inputs the normalized cross-correlation function ($R_c$) and the ILD of the left and right input signals. Specifically, $R_c$ is computed as:
\begin{equation*} \label{eq_ccf}
R_c(\tau)=\frac{\sum_{m}{( x_l(m)-\overline{ x}_l )} \sum_{m}{( x_r(m-\tau)-\overline{ x}_r )}}
{\sqrt{ \sum_{m}{( x_l(m)-\overline{ x}_l )^2}}
\sqrt{ \sum_{m}{( x_r(m-\tau)-\overline{ x}_r )^2}}
},
\end{equation*}
where $x_l$ and $x_r$ are the acoustic signals at the left and right ears, $\overline{x}_l $ and $\overline{x}_r$ are average values of the signals over a window of size $2\tau_{max}$, m is the sample index, $\tau \in [-\tau_{max}, \tau_{max}]$ is delay in time, and $\tau_{max}$ is the maximum delay that a normal human can perceive, about 1ms. For sounds sampled at 44.1kHz, $\tau_{max}$ corresponds to 45 samples. Therefore, this feature has a dimension of 91.  The ILD of $x_l$ and $x_r$  is a scalar defined as:
\begin{equation*} \label{eq_ild}
\begin{split}
\text{\emph{ILD}}=10\log_{10}\frac{\sum_{m}{x^2_l(m)}}
{ \sum_{m}{x^2_r(m)}}.
\end{split}
\end{equation*}
By concatenating the two, a feature vector of length 92 is input to the neural network. Since the model can only predict azimuth angles, the location error is defined as:
\begin{equation*} \label{eq_sl_err}
\begin{split}
\text{\emph{Error}}=\frac{\sum_{n=1}^{N}{\lvert\theta(n)-\hat{\theta}(n)\rvert}}{N},
\end{split}
\end{equation*}
which is an average over the errors between the ground truth ($\theta$) and estimated ($\hat{\theta}$) azimuth angles over $n$ test locations.

\begin{figure}[!t]
    \centering
    \includegraphics[width=2.1in]{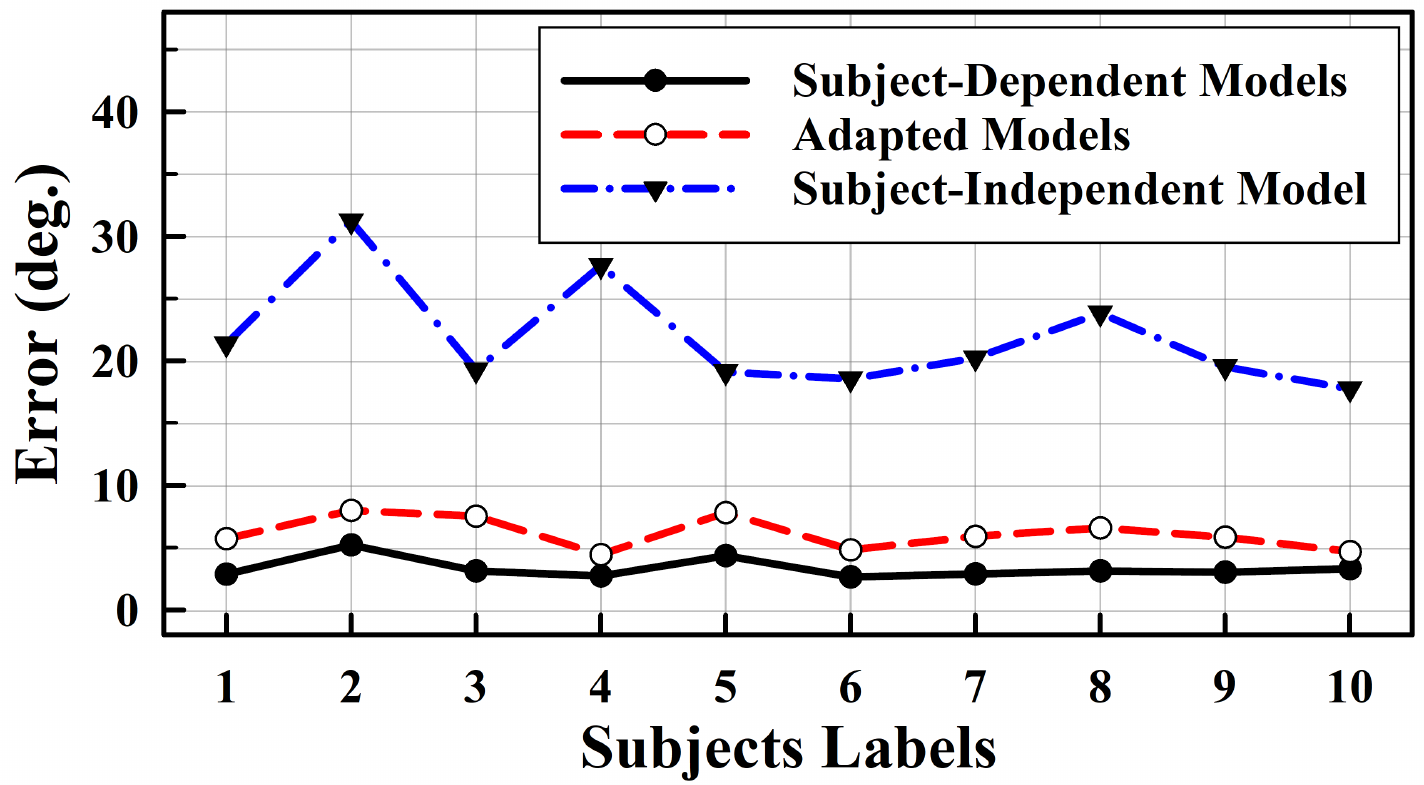}
    \caption{Angle estimation error of the subject-independent model, the subject-dependent and the adapted models for different subjects.}
    \label{table:sl_performance}
\vspace{-1em}
\end{figure}

For comparison, we train three types of models, namely, a \emph{subject-independent model}, \emph{subject-dependent models} and \emph{adapted models}. These models follow the same architecture as the base model but differ in model parameters due to different training data used. For training and testing, we take recordings of different types of sounds from the Harvard Sentences dataset \cite{S62, S63} and convolve them with selected HRTFs at target directions to generate binaural sounds.
The subject-independent model is trained using data from HRTFs of all subjects in the ITA dataset. It captures the {\it average} model of different subjects, and thus serves as a baseline. Ten subject-dependent models are trained using binaural sounds generated from known HRTFs of the respective subjects in the ITA dataset. Finally, for each of the 10 subjects, we first obtain its individualized HRTFs following the algorithm in Section~\ref{sec:individualization_decoder} using sparse samples at 80 locations in the frontal semi-sphere. The resulting training data is then used to update the parameters of the baseline model to produce the adapted model of each subject.

\emph{Results}. The average errors of the estimated azimuth of the subject-independent, subject-dependent, and adapted models are shown in Figure \ref{table:sl_performance}. The results are the averages of 6912 testing locations for each test subject. From the figure, we observe that the adapted models have a comparable accuracy to the subject-dependent models and outperform the subject-independent model. The classification accuracy of the subject-dependent, adapted, and subject-independent models is $96\%$, $92\%$, and $74\%$, respectively. The average errors for the subject-dependent and the adapted models are $3.41^\circ$ and $6.19^\circ$, respectively, while the average error is $21.92^\circ$ for the subject-independent model.

\begin{figure}[!t]
    \centering
    \includegraphics[width=2in]{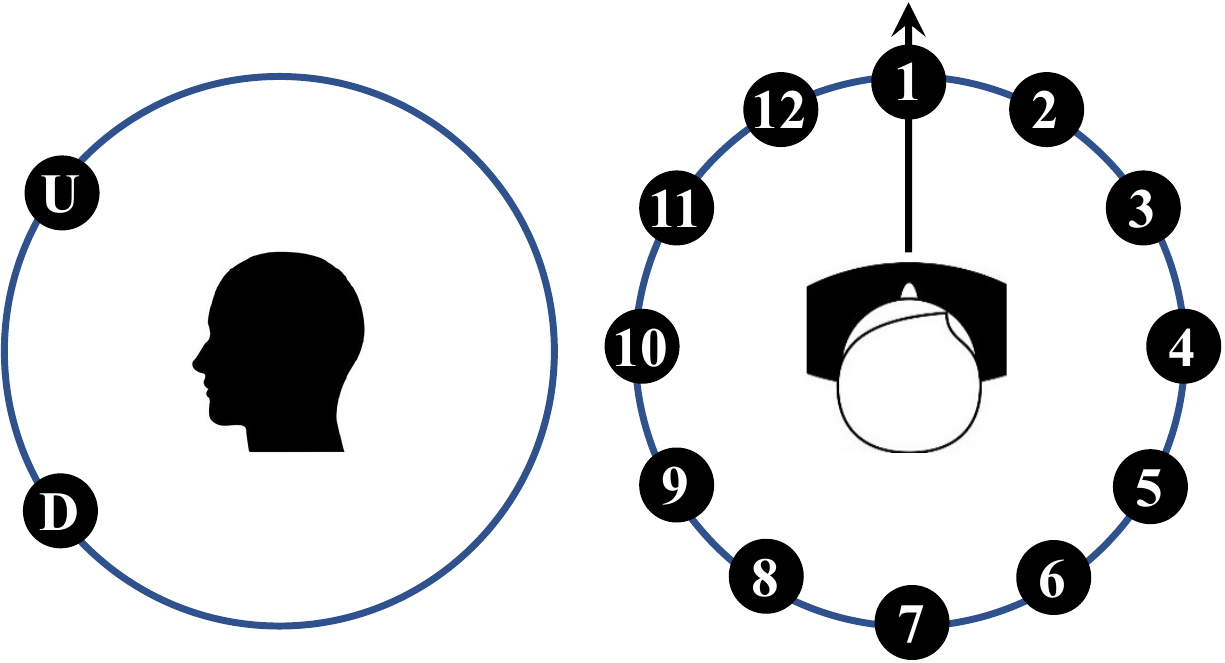}
    \caption{A questionnaire with 12-azimuth and 2-elevation candidate locations. A user is asked to select one from  each chart in spatialization tests.}
    \label{fig:questionnaire}
\vspace{-1em}
\end{figure}

\subsection{Acoustic Spatialization \label{Acoustic_spat}}
Acoustic spatialization is another application that can benefit from individualized HRTFs. Extensive researches have been done in the area of 3D sound spatialization. Some of its applications include interactive virtual environments to foster cognitive and learning skills in blind children \cite{sanchez20063d}, or as a portable navigation system for the blinds \cite{aguerrevere2004portable}. Also it is used in VR and 3D games, as an extra navigational aid to interact better with complex environments \cite{gunther2004using}, and to simulate moving sounds \cite{keyrouz2006rational}. At home children schooling is another new application that uses binaural hearing \cite{binauricsaudio}. Children can attend lectures using their tablet, as if they are in the middle of the classroom. All the mentioned applications are taking advantage of HRTF for the 3D sound spatialization. However, without an individualized HRTF, the user will not feel comfortable using them. Therefore, we perform this experiment to collect data from the subjects by measuring their HRTFs at sparse locations as detailed in Section \ref{experimental_valid}, and train subject-dependent decoders to generate their respective HRTFs in different directions.

For each subject, 14 sound files have been prepared by convoluting a mono sound (e.g., a short piece of music) with individualized HRTFs at directions chosen randomly from 12 azimuth angles evenly distributed between $0^\circ$ and $360^\circ$, and two elevation angles (Figure \ref{fig:questionnaire}). Additionally, we also prepare 10 sound files by convoluting the same sound with HRTFs of an arbitrary subject in the ITA dataset at different azimuth and elevation angles. The two sets of sound files are then mixed and shuffled. The subject is then asked to play back all sounds using a headset and label their perceived sound locations among the possible azimuth and  elevation angles.

The procedure is repeated for all subjects in Section \ref{experimental_valid}. We find that with the individualized HRTFs, subjects can accurately detect the correct azimuth angles $82.55\%$ of the time on average. The accuracy drops to a mere $29.17\%$ of the time when the unmatched HRTFs are used. However, in both cases, all subjects report difficulties in determining elevation angles. This is consistent with the fact that human auditory systems generally have poor elevation resolution \cite{human_factorsEle}. Therefore, we conclude that HRTF individualization can indeed provide more accurate acoustic spatialization and thus better 3D immersion experiences to users.

\section{Conclusion}
A new approach to HRTF individualization was introduced in this paper, which with the help of a Generative Neural Network, can estimate the HRTF of a subject, using only sparse data from the subjects. A fast data collection procedure is devised that can be performed by users at home without specialized equipment. This proposed approach shows great improvements in adaptation time, compared to perceptual-based methods. Accuracy of the proposed approach has been investigated using both a public dataset and real-world measurements. The advantages of individual HRTFs have been demonstrated using binaural localization and acoustic spatialization applications.

One limitation of the data acquisition procedure is that subjects can not measure any points in the back semi-sphere, which lowers the accuracy of individualized HRTFs using only data from frontal semi-sphere. Another limitation arises from the fact that the measured HRTFs contain artifacts from room reverberations and the used speaker and microphone. \emph{As future work}, we will further develop faster and more robust data acquisition procedure that decouples environmental effects from subject-dependent HRTFs, and can be conducted in a continuous manner.

\bibliographystyle{ACM-Reference-Format}
\bibliography{IEEEabrv, main}

\end{document}

%% file: direction_finding.tex
At each position, IMU sensor data of the mobile phone is logged to calculate the orientation of the phone in space.
Many sensor fusion algorithms can be utilized for this purpose such as the Mahony filter \cite{mahony2008nonlinear} and the Madgwick filter\cite{madgwick2010efficient}, both with the ability to mitigate magnetic interference from surrounding environments. However, the resulting phone orientation is with respect to a global coordinate frame (GCF). To determine the direction of sound sources, further transformations are needed to find the phone's azimuth and elevation angles in a head centered coordinate frame (HCF). Next, we describe the algorithm to do so. 

\subsubsection{Notations}
We first define the notations used in the algorithm as illustrated in Figure~\ref{fig:body_}. 
\begin{itemize}
\item Head-centered coordinate frame (HCF): a coordinate frame whose origin is at the centre of the head between a subject's two ears. Its $y$- and $x$-axes are both in a horizontal plane pointing to the front and right sides of the subject's body, respectively. The $z$-axis is vertical pointing upward,
\item Global coordinate frames (GCF): a coordinate frame centered on the shoulder joint of the phone holding hand with $y$- and $x$-axes pointing the geographical North and East, respectively. Its $z$-axis is vertical pointing away from the center of the earth. By default, we consider the GCF centered on the right shoulder joint  unless otherwise specified. 
\item $\alpha$: the rotation angle around $z$-axis from GCF to HCF clockwise.
\item $\phi_{m}$ and $\theta_m$ are respectively, the azimuth (with respect to the geographical North) and elevation angles of the phone's long edge (aligned with the subject's arm) in the GCF.
\item $\phi_{m}'$ and $\theta_m'$ are respectively, the azimuth and elevation angles of the phone's long edge (or equivalent the sound source) in the HCF.
\end{itemize}
and the anthropometric parameters of the subject are
\begin{itemize}
\item $l_{sh}$, the shoulder length of the subject from her left or right shoulder joint to the centre of her head,
\item  $l_s$, the distance from the subject's left and right shoulder joint to the phone speaker,
\item $l_z$, the vertical distance between the centre of shoulders, and the centre of the head.
\end{itemize}

Consider a point $P$ in space, whose coordinates in HCF and GCF are respectively, $(x', y', z')$ and $(x, y, z)$. From the above definitions, we note that GCF and HCF can be related by translations on $x$- and $z$-axes by $l_{sh}$ and $l_z$ and a rotation around the $z$-axis clockwise of an angle $\alpha$. Specifically, 
\begin{equation}
    \left[\begin{array}{c} x'\\ y' \\ z'\end{array}\right] = \underbrace{\left[\begin{array}{ccc}   \cos\alpha & -\sin\alpha & 0 \\
                                                \sin\alpha & \cos\alpha  & 0 \\
                                                0          & 0           & 1\\
                            \end{array}\right]}_{R_z(\alpha)}\left[\begin{array}{c} x\\ y \\ z\end{array}\right]  + \left[\begin{array}{c} l_{sh}\\ 0 \\ -l_z\end{array}\right],
\label{eq:translation_rotation}
\end{equation}
where $R_z(\alpha)$ is a rotation matrix around $z$-axis. 

\subsubsection{Algorithm}
When the phone is at azimuth  $\phi_m$ and elevation angle $\theta_m$ in the GCF, its Cartesian coordinates are $(l_s \cos\theta_m\sin\phi_m,$ $l_s \cos\theta_m \cos\phi_m, l_s \sin\theta_m)$. From \eqref{eq:translation_rotation}, its Cartesian coordinates in the HCF are expressed by \eqref{eq: HCF}. The azimuth and elevation angles of the sound source in the HCF are given by \eqref{eq:phi_m_prime} and \eqref{eq:theta_m_prime}, respectively. In \eqref{eq:phi_m_prime} and \eqref{eq:theta_m_prime}, the unknown parameters are $\alpha$, $\frac{l_{sh}}{l_s}$, and $\frac{l_z}{l_s}$. Note that there is no need to know the exact values of $l_{sh}$, $l_s$ and $l_z$, instead, the ratios suffice. Next, we detail the steps to determine these parameters {\it without} knowledge of anthropometric parameters.

\begin{figure}[!t]
\centering
\subfloat[The reference horizontal angle at ITD $=0$.]{\includegraphics[width=1.4in]{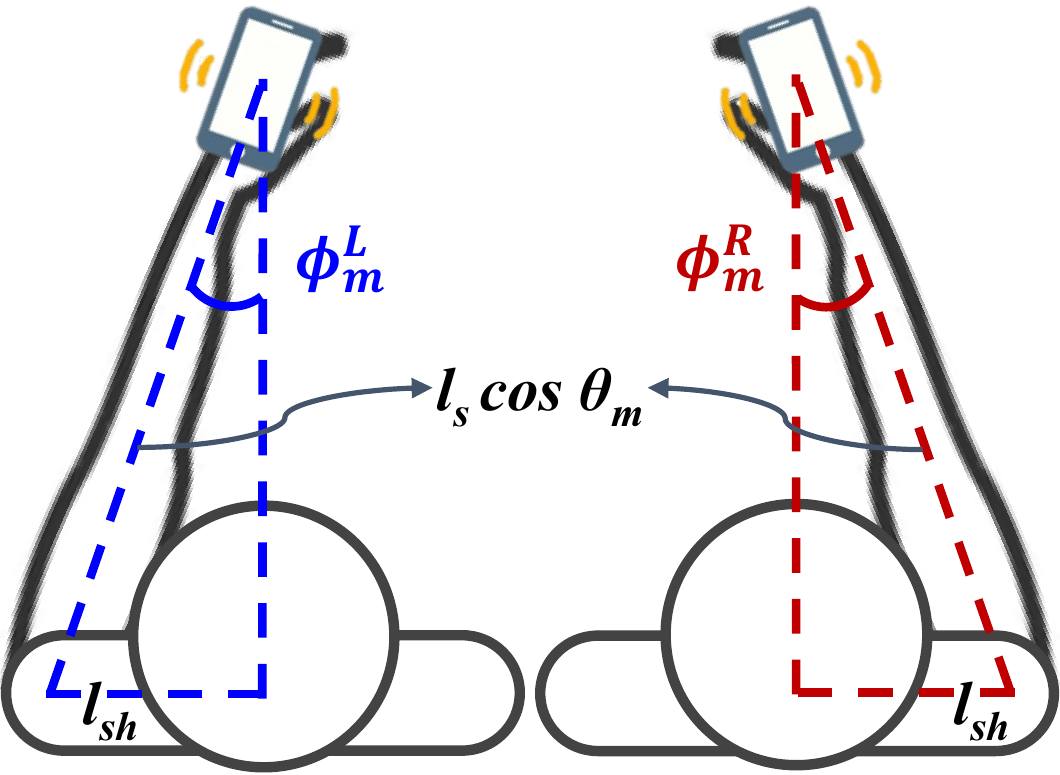}%
\label{fig:body_phi1}}
\hfil
\subfloat[Geometric laws that can be used to determine $l_{sh}/l_s$.]{\includegraphics[width=1.4in]{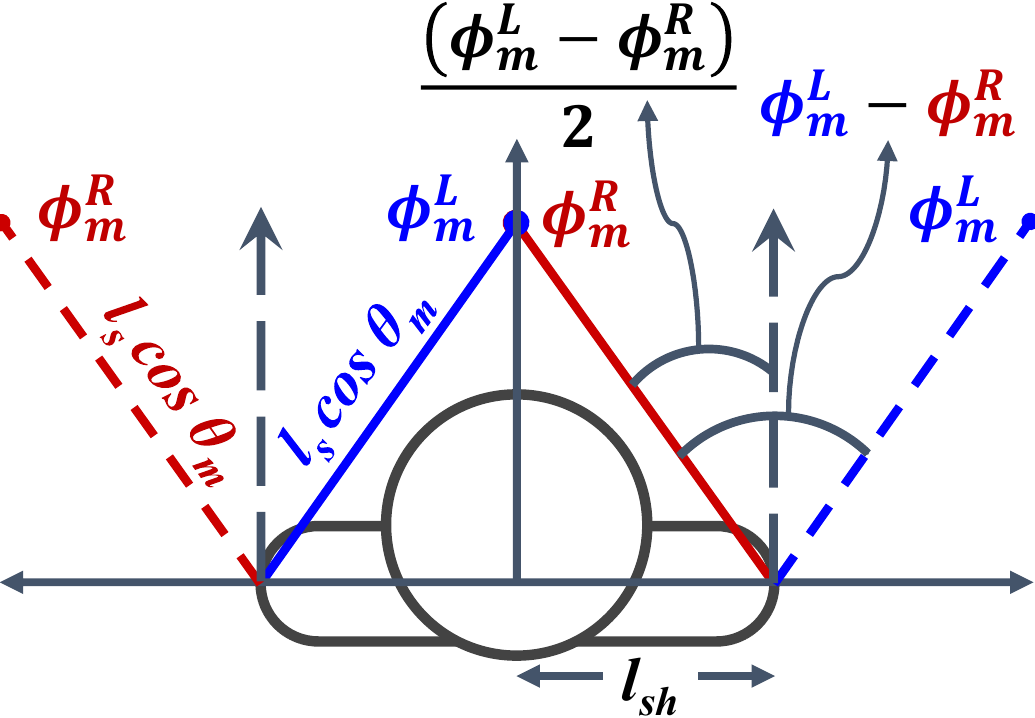}%
\label{fig:body_phi2}}
\hfil
\subfloat[Frontal and top views of reference angle associated with $\text{ITD}_{\text{max}}$.]{\includegraphics[width=2.8in]{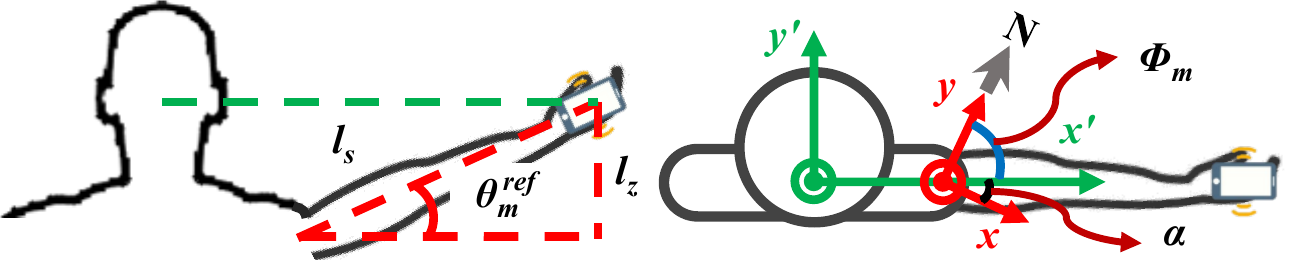}%
\label{fig:body_theta2}}
\caption{Geometrical relations in the horizontal and frontal planes. Measurements are done using both right and left hands. Reference angles can be found when ITD = 0 and $\mid$ITD$\mid$ reaches its maximum.}
\label{fig:body_phi}
\vspace{-1em}
\end{figure}

\begin{table*}[!b]
\hrule
\centering
\footnotesize
\begin{minipage}{\textwidth}
\begin{equation}
(l_s \cos\theta_m \sin\phi_m \cos\alpha - l_s \cos\theta_m \cos\phi_m \sin\alpha + l_{sh},~l_s \cos\theta_m \sin\phi_m\sin\alpha + l_s \cos\theta_m \cos\phi_m\cos\alpha,~l_s \sin\theta_m - l_z)
\label{eq: HCF}
\end{equation}
\begin{equation}
\phi_m' = \tan^{-1}\left(\frac{l_s \cos\theta_m \sin\phi_m \cos\alpha - l_s \cos\theta_m \cos\phi_m \sin\alpha + l_{sh}}{l_s \cos\theta_m \sin\phi_m \sin\alpha + l_s \cos\theta_m \cos\phi_m \cos\alpha}\right) = \tan^{-1}\left(\frac{\cos\theta_m \sin\phi_m \cos\alpha - \cos\theta_m \cos\phi_m \sin\alpha + \frac{l_{sh}}{l_s}}{\cos\theta_m \sin\phi_m \sin\alpha + \cos\theta_m \cos\phi_m \cos\alpha}\right)
\label{eq:phi_m_prime}
\end{equation}
\begin{equation}
\theta_m' = \tan^{-1}\left(\frac{\sin\theta_m - \frac{l_z}{l_s}}{\sqrt{(\cos\theta_m \sin\phi_m \cos\alpha - \cos\theta_m \cos\phi_m \sin\alpha + \frac{l_{sh}}{l_s})^2 + (\cos\theta_m \sin\phi_m \sin\alpha + \cos\theta_m \cos\phi_m \cos\alpha)^2}}\right)
\label{eq:theta_m_prime}
\end{equation}
\end{minipage}
\end{table*}

\emph{Estimating} $\frac{l_{sh}}{l_s}$. The key insight is, there are locations of the mobile phone associated with {\it known} azimuth or elevation angles in the GCF based on ITD measurements. Consider the positions of the phone in Figure~\ref{fig:body_phi1}. When the phone is on the sagittal plane that bisects the subject's body, the ITD to the left and right ears shall be zero. Let the corresponding azimuth angles of the phone held in the left and right hand be $\phi_m^L$ and $\phi_m^R$. From simple geometric relationships, we have $\frac{l_{sh}}{l_s} = \sin\left(\frac{\phi_m^L - \phi_m^R}{2}\right)\cos\theta_m$ as illustrated in Figure \ref{fig:body_phi2}. In practice, it is difficult for a subject to precisely place the phone in the sagittal plane. We can  approximate such locations by interpolating locations with small ITDs  when the phone is moved by both hands. 

\emph{Estimating} $\alpha$. When the phone is on the line connecting the subject's ears, the absolute value of ITD is maximized. Once such a position is identified (directly or via interpolation), we can estimate $\alpha$ as $\pi/2 - \phi_m$, as illustrated in Figure~\ref{fig:body_theta2}.

\emph{Estimating} $\frac{l_z}{l_s}$. Similarly, when the absolute value of ITD is maximized, we have $\frac{l_z}{l_s} = \sin\theta_m^{ref}$ (Figure~\ref{fig:body_theta2}).

To this end, we can estimate the three unknown parameters using only azimuth and elevation angles of the mobile device in the GCF and ITD measurements. At any position, given $\phi_m$ and $\theta_m$, we can then determine $\phi_m'$ and $\theta_m'$ using \eqref{eq:phi_m_prime} and \eqref{eq:theta_m_prime}.